\documentclass{aa}
\usepackage{graphics,graphicx}
\usepackage{epsfig,psfig}

\newcommand{\kms}{km\,s$^{-1}$}
\newcommand{\co}{$^{12}$CO J=2$\to$1$\,$~}
\newcommand{\cod}{$^{13}$CO J=1$\to$0$\,$~}

\begin{document}

   \title{Star formation in RCW~108: triggered or spontaneous?
\thanks{Based on observations obtained at the European Southern Observatory
using the ESO New Technology Telescope (NTT) (programs 61.D-0447,
64.L-0049, and 71.C-0429), the ESO-MPI 2.2m telescope (program
62.I-0454), the ESO 3.6m telescope (program 69.C-0522(A))
and the Swedish-ESO Submillimeter Telescope (program 61.C-0243), on La Silla, Chile.}}

   \author{F. Comer\'on\inst{1}
        \and
       N. Schneider\inst{2,3}
        \and
       D. Russeil\inst{4}
          }

   \offprints{F. Comer\'on}

   \institute{European Southern Observatory, Karl-Schwarzschild-Strasse 2,
              D-85748 Garching, Germany\\
          email: fcomeron@eso.org
          \and
           I. Physikalische Institut, Universit\"at K\"oln, D-50937 K\"oln, Germany
           \and
           OASU/Observatoire de Bordeaux, Universit\'e de Bordeaux I, F-33270 Floirac
           Cedex, France\\
           email: schneider@obs.u-bordeaux1.fr
         \and
           OAMP, 2 place LeVerrier, F-13004 Marseille, France\\
           email: delphine.russeil@oamp.fr
             }

   \date{Received; accepted}

\abstract{ We present visible, near infrared, and mm-wave
observations of RCW~108, a molecular cloud complex in the Ara~OB1
association that is being eroded by the energetic radiation of two
O-type stars in the nearby cluster NGC~6193. The western part of
the RCW108 molecular cloud, for which we derive a mass of $\sim
8000$~M$_\odot$, contains an embedded compact HII region,
IRAS~16362-4845, ionized by an aggregate of early-type stars for
which we estimate a mass of $\sim 210$~M$_\odot$. The spectral
type of the earliest star is O9, as confirmed by the visible
spectrum of the compact HII region. We notice a lack of stars
later than A0 in the aggregate, at least having the moderate
reddenings that are common among its B-type stars, and we
speculate that this might be a consequence of the extreme youth of
the aggregate. We also note the existence of a dense ionized clump
($n > 10^4$~cm$^{-3}$) appearing near the main ionizing star of
the compact HII region. We examine the distribution of stars
displaying infrared excesses projected across the molecular cloud.
While many of them are located in the densest ($n \sim 10^{4-5}$
cm$^{-3}$) area of the molecular cloud near the position of
IRAS~16362-4845, we also find a group concentrating towards the
edge of the cloud that faces NGC~6193, as well as some other stars
beyond the edge of the molecular cloud. The intense ionizing
radiation field by the O stars in NGC~6193 is a clear candidate
trigger of star formation in the molecular cloud, and we suggest
that the existence and arrangement of stars in this region of the
molecular cloud supports a scenario in which their formation may
be a consequence of this. However, infrared excess stars are also
present in some areas of the opposite side of the cloud, where no
obvious candidate external trigger is identified. The existence of
such tracers of recent star formation scattered across the more
massive molecular cloud associated with IRAS~16362-4845, and the
low star formation efficiency that we derive, indicate that it is
in a state to still form stars. This is in contrast to the less
massive cloud ($\sim 660$~M$_\odot$) close to NGC~6193, which
seems to be more evolved and mostly already recycled into stars,
and whose internal kinematics show hints of having been perturbed
by the presence of the massive stars formed out of it.\keywords
ISM: HII regions -- ISM: individual object: RCW~108 -- ISM: clouds
-- ISM: infrared }

\authorrunning{F. Comer\'on, N. Schneider, and D. Russeil}
\titlerunning{Star formation in RCW~108}
   \maketitle


\section{Introduction}

  \object{RCW~108} (Rodgers et al. \cite{rodgers60}) is an extended
HII region belonging to the \object{Ara~OB1} association (Herbst
\& Havlen~\cite{herbst77}, Kaltcheva \&
Georgiev~\cite{kaltcheva92}), highlighting one of the few
remaining places of Ara~OB1 where star formation is still active
(Yamaguchi et al.~\cite{yamaguchi99}). The existence and
arrangement of objects and structures usually associated to
massive star formation (early-type stars, molecular clouds with
dense cores, embedded pre-main sequence stars, and ionized gas)
makes it an interesting target for the study of the interplay
between the interstellar gas and dust and newly formed stars. Its
distance of approximately 1.3~kpc (see Arnal et
al.~(\cite{arnal03}) for a brief discussion; this is also the
value that we adopt in the present paper) allows a study at good
spatial resolution (1$'$ equals $\sim$0.4 pc). The HII region,
\object{NGC~6188}, is actually the bright rim of a molecular cloud
containing several star formation sites. The eastern edge of this
cloud is being eroded by the ionizing radiation of the nearby
O-type stars \object{HD~150135} and \object{HD~150136} in the open
cluster \object{NGC~6193}. Embedded in this cloud lies a compact
HII region (Shaver \& Goss~\cite{shaver70}) coincident with the
source \object{IRAS~16362-4845}. This region is easily noticed in
visible images of the area, but it is most prominent at near
infrared wavelengths at which the diffuse bright-rimmed HII region
is hardly seen. The study of the ionizing stars of the compact HII
region is especially interesting in the context of recent ideas on
the determinant role that Trapezium-like clusters may play on the
birth of high mass stars (e.g. Bonnell et al.~\cite{bonnell01},
Bonnell \& Bate~\cite{bonnell02}).

  A pioneering study of IRAS~16362-4845 was published by Straw et al.
(\cite{straw87}), who presented infrared imaging observations
ranging from the $J$ band (1.25~$\mu$m) up to 100~$\mu$m, as well
as low resolution infrared spectroscopy of two selected sources
and of the ionized nebula. Although that work established some of
the main characteristics of the compact HII region and its
associated stellar population, its depth and spatial resolution
are rather modest by the standards of current instrumentation on
medium-sized telescopes. Moreover, only IRAS~16362-4845 and its
immediate vicinity are considered in Straw et
al.~(\cite{straw87}). A recent study based on radio recombination
line, radio continuum, molecular line, mid-infrared (MSX), and
near-infrared (2MASS) observations of a region comparable to the
one discussed in the present paper has been published by Urquhart
et al.~(\cite{urquhart04}). Finally, low angular resolution
molecular observations of RCW~108 in the \cod line (Yamaguchi et
al.~\cite{yamaguchi99}) and the $^{12}$CO J=1$\to$0 line (Arnal et
al.~\cite{arnal03}) provide useful complementary information on
the larger scale distribution of lower density gas in the region
and the existence of other star forming sites in Ara~OB1.

This paper presents new visible and near infrared imaging and
low resolution spectroscopy of IRAS~16362-4845 and its surroundings,
aiming at complementing and updating the analysis of Straw et
al.~(\cite{straw87}) and at extending it to a broader area of the
molecular cloud. We also present maps in the lines of \co and \cod
centered on IRAS~16362-4845 covering most of the area included in our
near infrared images, and south of the emerged cluster NGC~6193.
Additional interferometric observations of the H$\alpha$ emission throughout
much of the region centered on IRAS 16362-4845 are shown and discussed.
Our imaging and molecular-line observations thus cover the densest
areas of the molecular clouds, whereas the low resolution spectroscopy
focuses on the compact HII region and its embedded stellar component.

\section{Observations\label{observations}}

\subsection{Near infrared imaging}

  A near infrared mosaic in the $J$ (1.25~$\mu$m), $H$ (1.65~$\mu$m),
and $K_S$ (2.2~$\mu$m) bands, covering an area of $13' \times 13'$
on the sky, was obtained on the night of 22/23 February 2000 using
the SofI infrared spectrograph and array camera at the ESO New
Technology (NTT) telescope. The central $5' \times 5'$ area
containing IRAS~16362-4845 was imaged at a greater depth by means
of stack of 15 frames in each filter, each one containing 6
individual exposures of 2~sec coadded on the detector, obtained
with small telescope offsets in between. The peripheral area was
imaged by pointing the telescope at 32 regularly spaced positions
defining the sides of a square centered on IRAS~16362-4845, each
frame being in turn the result of 6 individual exposures of 2
seconds coadded on the detector. The spacing between consecutive
pointings in the peripheral area was 1', i.e., 1/5 of the field of
view of the camera. Each sky position was thus imaged by five
different pointings, except near the borders of the square
pattern. After flat fielding and dark subtraction, the sky frame
was constructed by median averaging the stack of the periphery
frames (which were found to be virtually devoid of nebulosity)
with clipping of the upper half of the pixel values. This sky
frame was then subtracted from each individual pointing in both
the periphery and the central area. All the $K_S$-band images were
then registered to construct the mosaic, using the positions of
stars common to consecutive images as references to compensate for
the telescope offsets. The mosaics in the other two filters were
constructed by individually registering each of their component
frames with the $K_S$-band mosaic, in order to avoid slight
relative scale distortions between filters due to accumulation of
small errors in the registering process.

  Sources were detected in our frames using DAOFIND
(Stetson~\cite{stetson87}). Relatively isolated unsaturated images
of bright stars were used to determine an approximate PSF needed
for the identification of point sources. Due to the crowdedness of
the field outside the areas of densest nebulosity and to our
interest in point sources only, photometry on the resulting mosaic
was performed by defining an undersized aperture at the position
of each detected star, measuring the flux inside it, and then
adding the rest of the flux in the PSF as given by the fit of a
circularly symmetric radial profile to each stellar image. This
procedure allowed both to remove the contamination to the
photometry to other stars located on the wings of the PSF, and to
adjust to the mildly variable image quality across the field of
view.

\subsection{Visible spectroscopy}

  Spectroscopy was carried out on the night of 1/2 April
2003 using EMMI, the visible imager and spectrograph at the NTT.
We chose a grism yielding a coverage of the $3800~{\rm \AA} <
\lambda < 9700~{\rm \AA}$ interval at a resolution of $\lambda /
\Delta\lambda = 570$ with the 1''0-wide slit that we used. The
slit was placed on the line joining the brightest star seen
projected on IRAS~16362-4845 in visible-light images and the
faint, very red star 4''0 to its Southwest that becomes the
brightest one in the near infrared $K$ band\footnote{These are
respectively denominated Star~8 and Star~12 in our discussion of
the stellar aggregate associated to IRAS~16362-4845 in
Section~\ref{aggregate}}. The effective slit length, 8'0, provided
us with a cut across the compact HII region but also included a
bright segment of the rim nebula. We obtained 3 separate spectra
of 10~minutes of exposure time each on the same telescope
position, and then coadded the resulting frames. Spectra of the
two stars were extracted using the APALL task on IRAF, defining
sections adjacent to the traces of the stars for the subtraction
of the sky and nebular emission together. Spectra of the nebula at
selected points were extracted as well by interactively defining
the apertures on the frames and defining an appropriate sky
aperture in a suitable area (see discussion in
Section~\ref{spec_nebulae}). Relative flux calibration was
performed using the spectrum of LTT~4364~(Hamuy et
al.~\cite{hamuy92}) as a reference. Wavelength calibration was
carried out by extracting the spectrum of a ThAr lamp obtained
with the same instrumental setup at the same aperture positions.

\subsection{Near infrared spectroscopy}

  Near infrared simultaneous low resolution spectroscopy in the $H$
and $K$ bands was obtained using SofI in spectroscopy mode on the
night of 2/3 April 2003. The grating used covered the 1.5~$\mu$m -
2.4~$\mu$m interval at a resolution $\lambda / \Delta \lambda =
590$ with our 1''0 slit. In this case the slit was
placed on the line joining the brightest near infrared source and
another source 3''5 to its Southwest that is similarly bright in
the infrared but redder in color, thus being undetected in our
visible images\footnote{These are respectively denominated Star~12
and Star~16 in our discussion of the stellar aggregate associated
to IRAS~16362-4845 in Section~\ref{aggregate}}. Four individual
spectra of 50~sec exposure each were obtained by slightly
offsetting the telescope between exposures along the direction of
the slit. Spectra were then extracted using a procedure analogous
to that followed for the visible spectra. Cancelation of telluric
features in the extracted spectra was achieved by observing the
G3V star HIP~81746 immediately after the observation of RCW~108 at
a very similar airmass, and relative flux calibration was carried
out by assuming that the overall shape of the spectrum of
HIP~81746 is well approximated by a black body at a temperature of
5700~K over the 1.5~$\mu$m - 2.4~$\mu$m interval. Wavelength
calibration in the infrared was performed by using the airglow OH
emission lines as a reference (Oliva \& Origlia~\cite{oliva92}).

\subsection{Additional imaging observations\label{additional_obs}}

  Besides the imaging and spectroscopy described in the previous
Sections, we obtained some additional visible imaging that is
helpful in illustrating the overall, large scale morphology of
RCW~108 as well as the more detailed structure of the
IRAS~16362-4845 HII region. For this purpose, a H$\alpha$ image of
the area was obtained using the Wide Field Imager (WFI) at the
ESO-MPI 2.2m telescope on La Silla on the night of 26/27~March
1999. The image was obtained by combining four individual
pointings of the telescope with offsets of approximately 2' in
order to cover the gaps between the individual CCD chips composing
the WFI detector array. The exposure time of each individual frame
was 300~sec. Additional short exposures through the $B$ and $V$
filters were also obtained, allowing their combination into a
color image that has been widely reproduced elsewhere (see e.g.
Collins Petersen~\cite{collins01}).

  We also obtained $UBVRI$ images of a smaller, $5'3 \times 5'3$
field centered on IRAS~16362-4845 using the SUSI2 visible imager
on the NTT on the night of 26/27 August 1998. Since the SUSI2
detector is also composed of two chips separated by a gap we
divided the exposures into three pointings separated by small
telescope offsets, allowing the reconstruction of a continuous
image of the field with the gaps filled. The total exposure time
in each filter is 9 minutes.

\subsection{Millimeter CO observations}

Observations of \co at 230.538~GHz and \cod at 110.201~GHz were
carried out using the SEST during September 1998 towards
IRAS~16362-4845 (an area of $\sim10' \times 8'$ was mapped) and
south of the cluster NGC 6193 ($4' \times 7'$). We refer hereafter
to the molecular gas detected at those positions as {\it the
western} and {\it the eastern} clouds, respectively. The map
centers were $\alpha(2000)$=$16^h40^m00^s13$,
$\delta(2000)$=$-48^\circ 51' 46.5''$ and
$\alpha(2000)$=$16^h41^m20^s32$, $\delta(2000)$=$-48^\circ
47'45.1''$ respectively. A pointing grid of 22 arcsec was adopted,
similar to the beamsize at 230~GHz (23$''$) and half-beamsize at
110 GHz (45$''$). The velocity resolution of all data is
0.11~\kms\ and the velocity coverage is $-$32 to $-$7~km~s$^{-1}$
for the eastern cloud (around NGC~6193) and $-$36 to
$-$13~km~s$^{-1}$ for the western cloud (around IRAS~16362-4845).
The average system temperature throughout the observations was
210~K for \co and 140 K for \cod .  The calibration was done using
the standard chopper-wheel method.  The resulting
atmosphere-corrected antenna temperatures were converted to
main-beam brightness temperatures using the values for main-beam
efficiencies quoted in the SEST Handbook (0.5 for 230~GHz and 0.70
for 110 GHz). The telescope pointing and subreflector focusing
were checked regularly, we estimate the pointing accuracy to be
better than 5~arcsec and adopt the standard SEST value of 10\% for
the uncertainty in the antenna temperature scale.  The average
r.m.s. main beam brightness temperature noise per channel is 0.5~K
for \co and 0.2~K for \cod .

\begin{figure*}
  \resizebox{18cm}{!}{\includegraphics{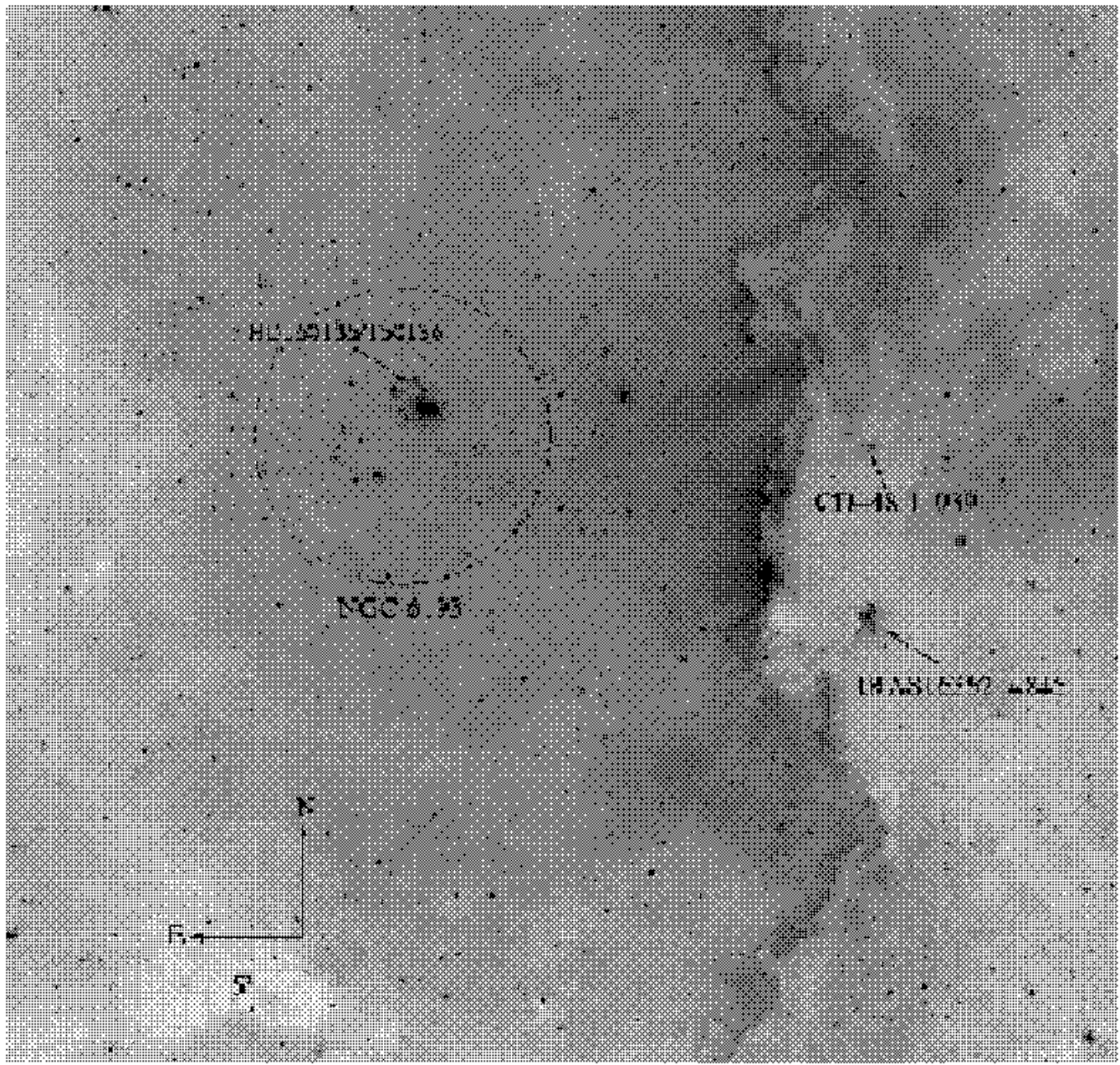}}
  \caption[]{A negative H$\alpha$ wide-field image of the RCW~108
area (Sect.~\ref{additional_obs}), showing the overall
distribution of the ionized gas in the region and main objects
discussed in the text. The large-scale distribution of the column
density of obscuring dust can be inferred as well by noting the
highly variable surface density of background stars across the
field. The rings appearing around or next to bright stars are
out-of-focus ghost images caused by internal reflections.}
  \label{wfi_halpha}
\end{figure*}

\subsection{Interferometric H$\alpha$ observations}

The H$\alpha$ interferometric observations were made with the
CIGALE instrument on the 3.6m telescope (La Silla) in April 2002.
The data cubes were obtained with a spatial resolution of 0.4$''$
for 4 fields centered on the compact HII region.  The size of each
field is 4.5 arcmin. The Fabry-Perot interferometer used has an
interference order 1938 (at H$\alpha$ wavelength) providing a
spectral sampling of 3.2~km~s$^{-1}$ and a free spectral range of
155~km~s$^{-1}$. The interference filter used is centered at
6562~\AA\ with a FWHM of 11~\AA.  The velocity and line width
accuracy is estimated to be 0.7~km~s$^{-1}$.  A complete
description of the instrument, including data acquisition and
reduction techniques has been given in le Coarer et
al.~(\cite{lecoarer92}).  The H$\alpha$ profiles are decomposed
with the H$\alpha$ geocoronal night-sky line (modeled by a purely
instrumental profile) and a nebular line (modeled by an
instrumental profile convolved with a gaussian).  An automatic
procedure has been used to perform the fit. Each field contains
512$\times$512 pixels, but in order to increase the
Signal-to-Noise ratio (S/N) we extracted and analysed profiles
from areas of 6.5$'' \times$6.5$''$ size.  A rough flux
calibration was performed by observing the planetary nebulae
NGC~2899 and adopting fluxes given by Perinotto and
Corradi~(\cite{peri98}).

\section{Results}

\subsection{Morphology of the distribution of gas\label{morphology}}

  The overall distribution of ionized and molecular gas over the
RCW~108 area and its surroundings is well illustrated by the
negative wide-field H$\alpha$ image presented in
Figure~\ref{wfi_halpha}. The cluster NGC~6193 occupies the
central/eastern side of the field and is dominated by the close
pair of O stars HD~150135/150136. Glow in H$\alpha$ {-- visible as
dark patches -- is present over most of the field, mostly arising
from gas on the background of NGC~6193, as well as probably from
gas left over from the formation of the cluster indicated by a
peak in molecular line emission seen in the same direction as its
brightest stars (see Section~\ref{molecular}). A simple inspection
of the starcount density in the area to the East of the rim nebula
indicates that the extinction on the background is low in the
Northern half and higher towards the Southeast.

  The arrangement of illuminated and shadowed areas in the rim
nebula suggests that most of the interface between the molecular
cloud and the ionization front eroding it is seen roughly edge-on.
The pattern of bright nebulosities and shadows dramatically
reveals the intricate three-dimensional structure, as well as the
existence of regions with widely varying densities. The column
density of the gas near the Northern and Southern edges of the rim
nebula is low, as seen from the only slight decrease of stellar
density in its direction, as compared to the more opaque clouds
located near the center of the image. Small, high density cores
are visible as dark patches projected against a brighter
background all over the nebula, and some of them appear on areas
where the erosion front has traveled past them, leaving them
isolated from the bulk of the molecular gas to the West.

The most conspicuous star forming site of RCW~108,
IRAS~16362-4845, roughly coincides with the thickest part of the
molecular cloud, as shown by both starcounts and molecular-line
maps (Section~\ref{molecular}; see also
Figure~\ref{moloverview})}. North of it one finds the reflection
nebula surrounding the B3V star \object{CD-48~11039}
(Herbst~\cite{herbst75}). No other star forming sites are obvious
in either visible or infrared images, although we will discuss
evidence for other lower-mass star forming sites spread across
RCW~108 in Section~\ref{other_sf}.

\begin{figure*}
  \resizebox{18cm}{!}{\includegraphics{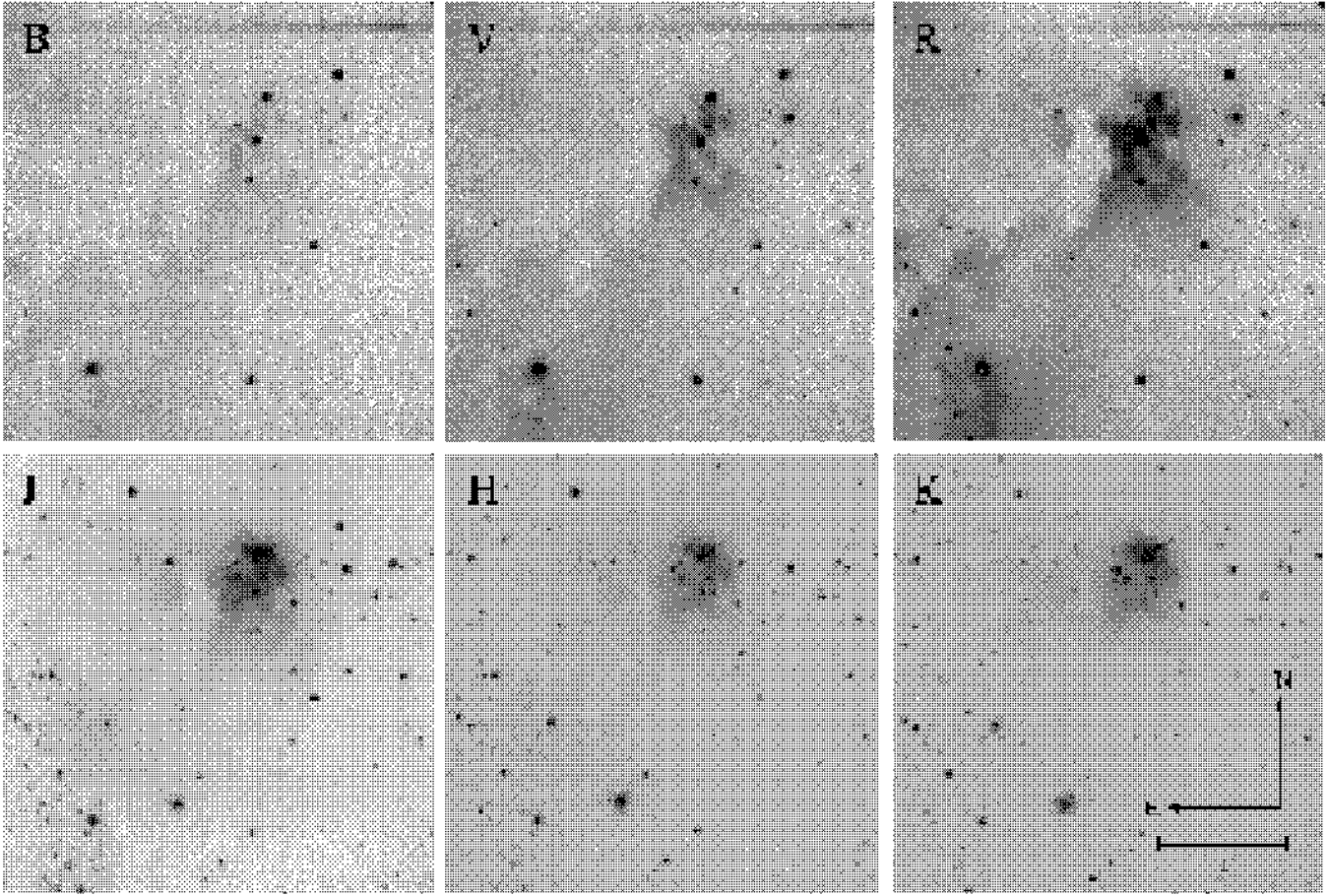}}
  \caption[]{A sequence of blue to infrared images of
IRAS16362-4845 field, showing the changing appearance of the
nebulosity as the extinction decreases when going to longer
wavelengths. Visible images show all-pervasive, mostly reflection
nebulosity, at least part of which is illuminated by stars not
related to the HII region. The latter gains prominence in the red
and infrared, revealing the aggregate of stars responsible for its
ionization.}
  \label{from_vis_to_ir}
\end{figure*}

  A comparison of visible and near infrared images of
IRAS~16362-4845 and its surroundings, shown in
Figure~\ref{from_vis_to_ir}, provides useful information on the
structure of the compact HII region. In the visible, the brightest
patch of nebulosity lies at the position of three bright and only
lightly reddened stars evenly spaced roughly in the North-South
direction. The peak of visible H$\alpha$ emission is displaced to
the South of the nominal IRAS position, where the visible
H$\alpha$ emission is comparatively faint. This indicates that the
core of the compact HII region is heavily reddened, and thus that
the northernmost of the three bright visible stars is not its
ionizing star, as already noted by Straw et al.~(\cite{straw87}).

  Broad-band visible images in $B$ and $V$ show reflection
nebulosity to the Southeast of the zone of most intense H$\alpha$
emission. This might be caused by a relatively unimpeded line of
sight from that vantage point to the stars ionizing the HII
region. However, we believe it more likely that the source of
illumination is actually an anonymous star located to the
Southeast, at $\alpha(2000) = 16^h40^m08^s21$, $\delta(2000) =
-48^\circ 53' 49''9$. Although this star does not show any
distinctive signs of belonging to the association from the data at
hand, indirect evidence comes from a very red source lying 28'' to
its Northeast, which seems to be still embedded in the remnants of
the core from which it formed. This core casts a shadow on the
surrounding nebula that points directly away from the star noted
above.

  The relatively unobscured view into the nebula provided by
the infrared images shows important differences with the
visible-light picture. The $K$-band image, which mainly traces the
emission in Br$\gamma$ (2.166~$\mu$m) and HeI (2.058~$\mu$m), is
now clearly peaked at the position of the IRAS source, where a
tight cluster of reddened stars (Section~\ref{aggregate}) is seen,
and extends eastwards in the general direction of NGC~6193. This
eastward extension is totally blocked from view in visible images
by an opaque layer of dust that in some small areas is thick enough
to block the background emission even in the $K$ band.

  To summarize, the overall structure of IRAS~16362-4845 hinted
by the visible and infrared images is that of an embedded compact
HII region. The foreground extinction decreases towards the South,
where the peak of the visible emission is, but the HII region
itself presents an extension towards the East that lies behind a
thicker layer of obscuring dust. At the adopted distance of
1300~pc the size of the compact core is $0.16$~pc ($25''$),
surrounded by a fainter halo of $0.22$~pc ($35''$) in diameter,
and the eastward extension reaches up to $\simeq 0.65$~pc
($100''$) from the center of the core. In the following section,
we present the results of our molecular line mapping which support
the scenario described above.

\begin{figure*}
\begin{center}
  \resizebox{13cm}{!}{\includegraphics{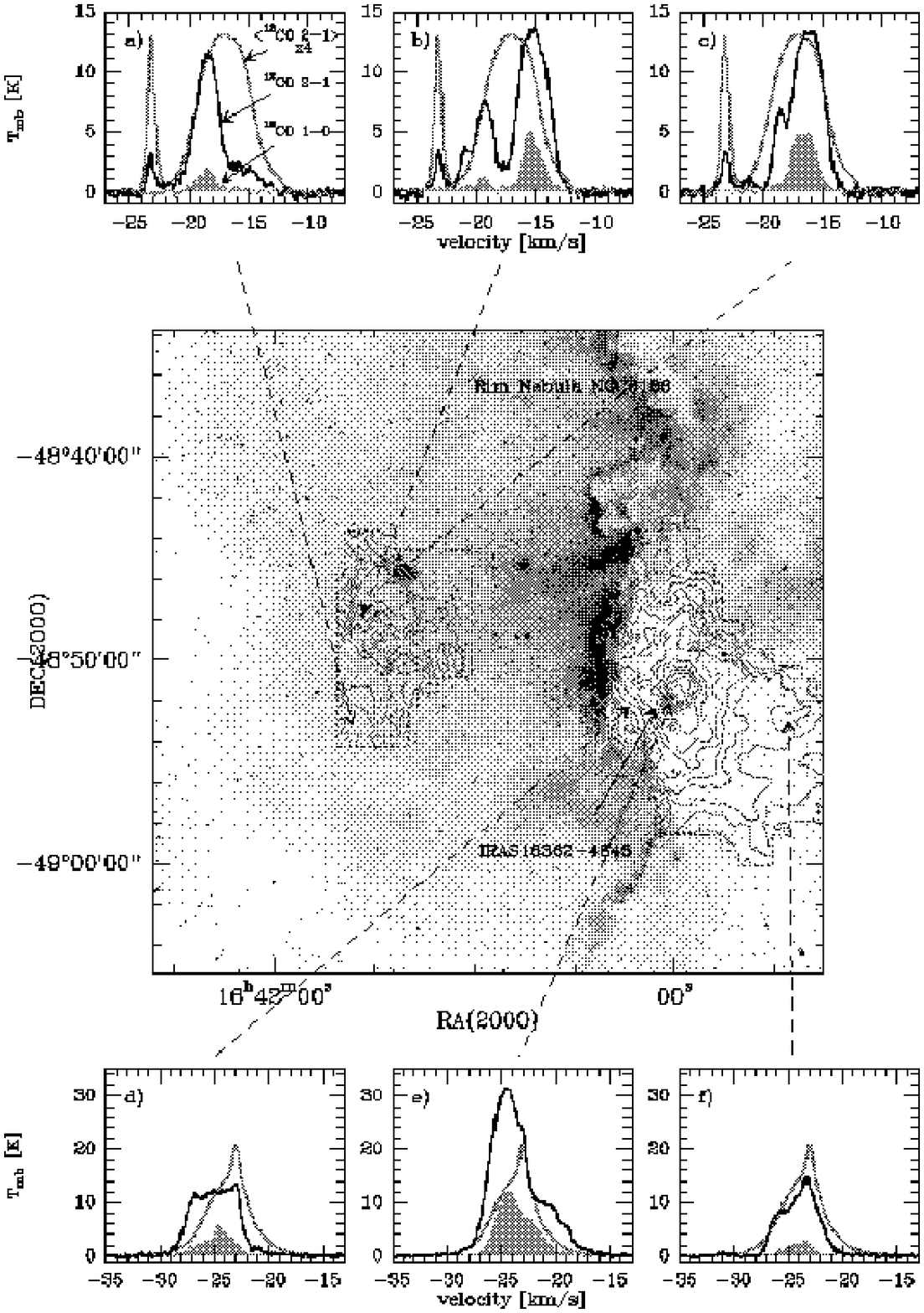}}
  \caption[]{The same negative H$\alpha$ wide-field image of RCW~108
as shown in Fig.~\ref{wfi_halpha} is overlaid with contours of
velocity integrated \co emission at 23$''$ angular resolution. The
velocity range for the eastern (western) cloud is $-$24 to
$-$12~km~s$^{-1}$ ($-$30 to $-$18~km~s$^{-1}$) and contours go from 17
(25) to 119 (300)~K~km~s$^{-1}$ in steps of 17 (25)~K~km~s$^{-1}$.
Representative \co (black line) and \cod (filled grey) spectra
taken at different positions from the eastern and western
molecular clouds are shown in the panels above and below. For each
cloud a positionally averaged \co spectrum (grey line), amplified
by a factor 4, is displayed as well. }
  \label{moloverview}
\end{center}
\end{figure*}

\begin{figure*}
\begin{center}
  \resizebox{13.5cm}{!}{\includegraphics[angle=-90]{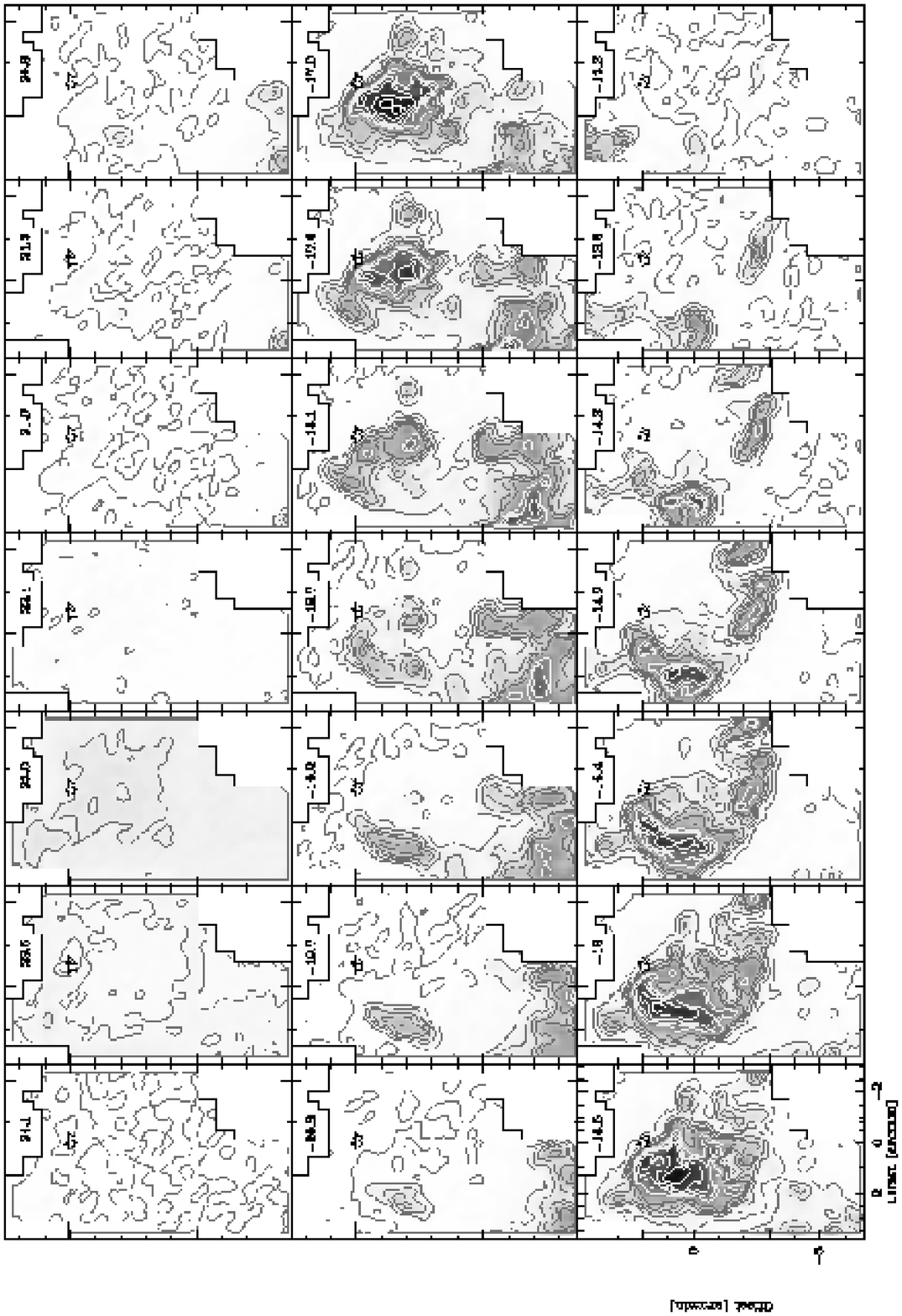}}
  \resizebox{13.5cm}{!}{\includegraphics[angle=-90]{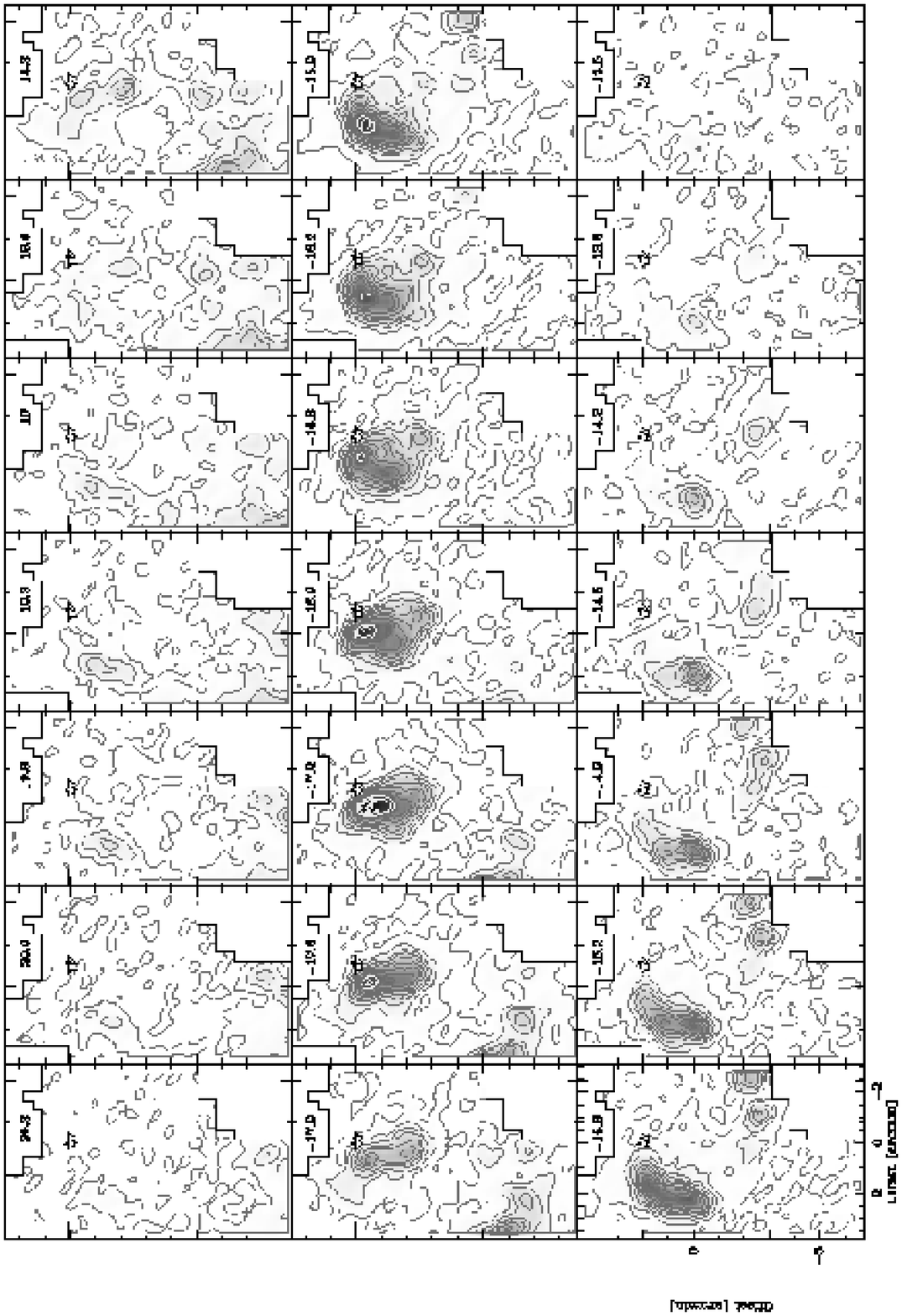}}
  \caption[]{
\co (top) and \cod (bottom) emission distribution around NGC 6193
in the velocity range $-$24.1 to $-$12.2~km~s$^{-1}$ ($-$20.3 to
$-$13.5~km~s$^{-1}$). The velocity coverage of each \co (\cod) map
is $\sim$0.5 (0.3)~km~s$^{-1}$ (five (three) velocity channels are
comprised) and the central velocity is indicated in each panel.
The stars represent HD 150135 and 150136.  Contours go from
3$\sigma$ (1.5~K~km~s$^{-1}$) to 453$\sigma$ in steps of
30$\sigma$ for \co and 3$\sigma$ (0.4~K~km~s$^{-1}$) to
101$\sigma$ in steps of 15$\sigma$ for \cod.}
  \label{chan_ngc}
\end{center}
\end{figure*}

\begin{figure*}
\begin{center}
  \resizebox{13.5cm}{!}{\includegraphics[angle=-90]{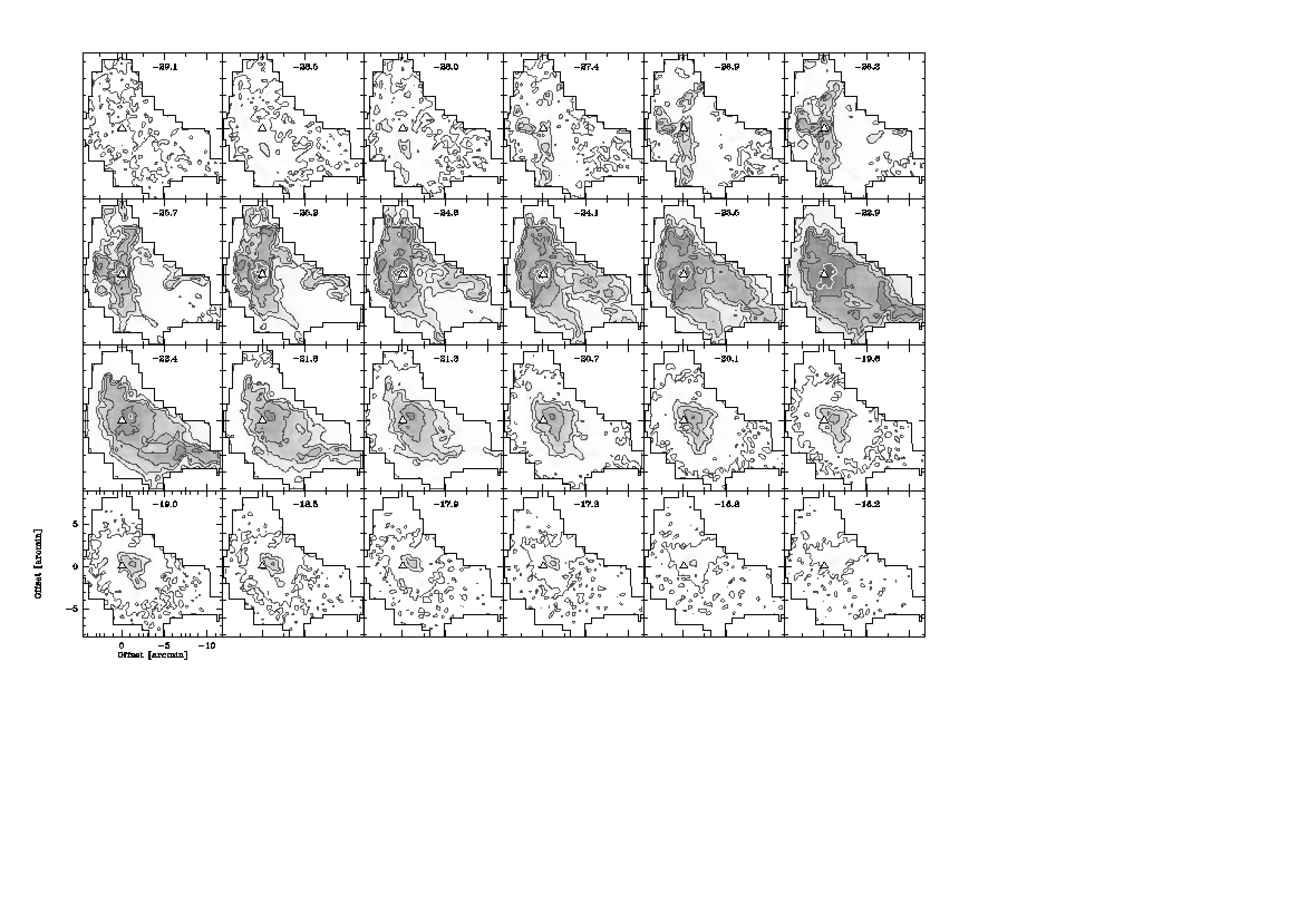}}
  \resizebox{13.5cm}{!}{\includegraphics[angle=-90]{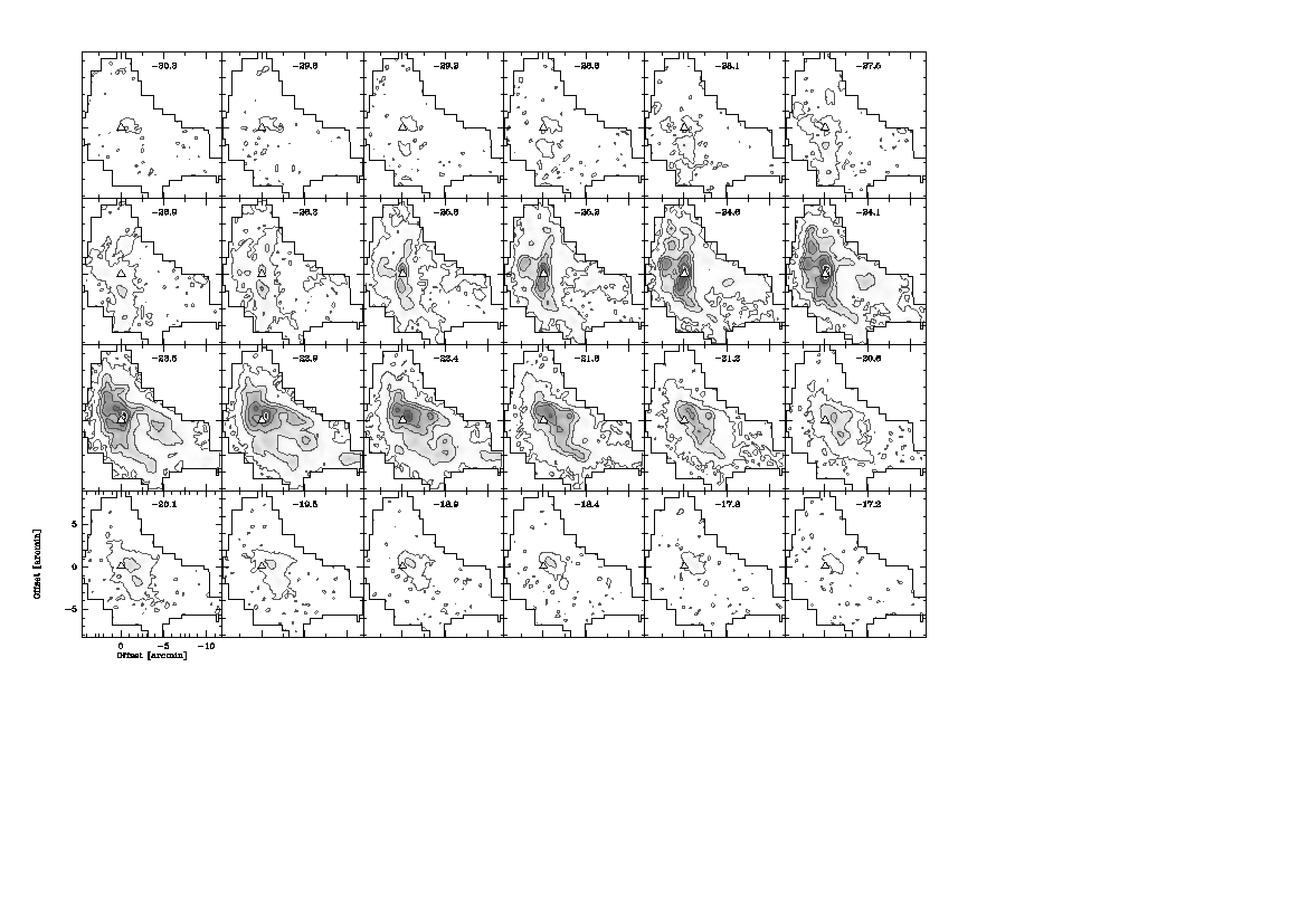}}
  \caption[]{
\co (top) and \cod (bottom) emission distribution around
IRAS~~16362-4845 in the velocity range $-$29.1 to $-$16.2 km
s$^{-1}$ ($-$30.3 to $-$17.2~km~s$^{-1}$). The velocity coverage
of each map is $\sim$0.5~km~s$^{-1}$ (five velocity channels are
comprised) and the central velocity is indicated in each panel.
Contours go from 3$\sigma$ (1.5~K~km~s$^{-1}$) to 453$\sigma$ in
steps of 45$\sigma$ for \co and 3$\sigma$ (0.4~K~km~s$^{-1}$) to
303$\sigma$ in steps of 30$\sigma$ for \cod. The triangle
indicates the position of IRAS~~16362-4845.}
  \label{chan_rcw}
\end{center}
\end{figure*}

\subsection{Molecular line maps\label{molecular}}

\subsubsection{Overview\label{overandspectra}}

  We focussed on mapping the environment of IRAS~16362-4845 and
NGC~6193 in the \co and \cod lines. The normally optically thin
\cod line is used to determine the column density and mass of the
molecular clouds (Sec.~\ref{masses}) while the \co line provides
an overview of the general small scale (23$''$) distribution of
molecular gas even on a low intensity level.
Figure~\ref{moloverview} (center) presents an overlay of \co
emission on the optical image where the eastern and western
molecular clouds apparently associated with IRAS~16362-4845 and
NGC~6193 are outlined by contours of CO emission. The western
cloud corresponds very well to a region of high extinction (white
areas in the H$\alpha$ image) which is not the case for the
eastern cloud fragment. There, regions of high extinction are more
diffuse and less clearly defined.

  Arnal et al.~(\cite{arnal03}) mapped a 2.75$^\circ
\times$3$^\circ$ region around RCW~108 at 8.$'$7 resolution in
$^{12}$CO 1$\to$0 and revealed a network of molecular clouds at
velocities between $-$50 and +6~km~s$^{-1}$.  However, by using
the galactic rotation curve they conclude that only CO emission
between $-$27 and $-$15~km~s$^{-1}$ is related to the Ara OB1
association. This velocity range matches to our maps and indeed,
we observe a very close correspondence between the western cloud
and features related to Ara OB1 (the rim nebula NGC~6188 and
IRAS~16362-4845): the interface between HII region and western
molecular cloud is impressively outlined by a sharp gradient of
molecular line emission and follows closely the optical features.
This indicates also that we see this region edge-on as already
concluded from the H$\alpha$ image alone
(Section~\ref{morphology}). Another prominent feature of the
western cloud is the marked peak of \co emission close to the
position of IRAS~16362-4845. The ionized gas of this compact HII
region is seen as a dark patch slightly shifted south-east from
the main CO peak. Considering the distortion of the CO contour
lines in this area, a close interaction between ionized and
molecular gas is likely, i.e. erosion of molecular clumps by
streaming ionized gas.

  In contrast to this rather evident correspondence between ionized
and molecular gas, the molecular emission close to the pair of O
stars HD 150135/150136 does not show a clear morphology which is
related to these stars. It shows a globular structure with a
region of high column density pointing away from the stars in
south-east direction. Lower density material without directed
structure is present in the south and southwest. This cloud was
mapped at 2$'$ angular resolution by Phillips et al.
(\cite{phillips86}) and they concluded that it represents a
wind-swept globule probably shaped by NGC~6193, i.e. HD
150135/150136. In our velocity integrated map, this scenario is
not so clear and we will come back to this point in the next
section where channel maps are discussed.

  A first impression of the complex velocity structure of the
molecular clouds is given by the \co and \cod spectra displayed in
Fig.~\ref{moloverview}.  The upper panels from the NGC 6193 region
show that there are several velocity components which are present
at all positions but vary in intensity. The most distinct feature
is a single line at $-$23~km~s$^{-1}$ whereas the other lines
between $-$21 and $-$12~km~s$^{-1}$ partly overlap. However, there
is a clear velocity gradient across the cloud, visible as a line
shift compared with the the average \co spectrum (in grey) from
$-$19~km~~$^{-1}$ (a) to $-$15~km~s$^{-1}$ (c). Since \co and \cod
show the same line profile, effects like self-absorption can be
excluded in first order.

The lower panels represent three positions from the western cloud
where the line profiles are less variable. The \co spectrum from
the interface region (d) displays a flat-top profile (while the
\cod line looks like a blending of several gaussian components),
probably indicating self-absorption effects at all velocities. In
contrast, the \co spectrum from a quiescent cloud region (f) and
the average profile show a decline in intensity only between $-$30
and $-$24~km~s$^{-1}$. The \cod spectrum from the peak position of
IRAS~16362-4845 (e) shows that at least two Gaussian lines --
though blended -- at --22 and --24~km~s$^{-1}$ can be identified.
We find at this position most intense \co brightness temperatures
of $\sim$30~K. Interestingly, the \co spectrum even indicates
additional emission in the form of a non-gaussian broad wing
between --21 and --17 km s$^{-1}$ which is not present in the
other spectra. Since this emission feature is not well spatially
focused (see Fig.~\ref{chan_rcw}) and only very weak wing emission
on the blue side of the spectrum (between --27 and
--29~km~s$^{-1}$) is found, we assume that it is probably not due
to the outflow emission of a YSO. As we will see in the channel
maps (Fig.~\ref{chan_rcw}), the --21 and --17 km s$^{-1}$ emission
feature is only found at the position of IRAS~16362-4845 and -
even more prominent - northwest of this source. It may represent a
second component of the molecular cloud showing a close
interaction with the HII region: we possibly observe swept-up
molecular material, arising from clumps eroded by ionizing gas
from the compact HII region. Since the peak of \co and \cod
emission is not found at the position of IRAS~16362-4845, the
double-peak \co emission features can probably not exclusively be
explained by a symmetric expanding shell where the compact HII
region created a cavity. This scenario is suggested by Urquhart et
al.~(\cite{urquhart04}) for explaining their CO observations at
the position of IRAS source. A more complicated geometry with an
embedded compact HII region and a dense molecular clump at the
north-western border of the cavity is probably more likely.


\begin{table*}
\caption{Excitation conditions of the molecular clouds associated
with IRAS~16362-4845 (western cloud) and NGC~6193 (eastern cloud).
Column 1: Cloud labeling. Cols. 2 and 3: average main beam
brightness temperatures of the \cod and \co lines. Col. 4: line
center velocity of \cod. Col. 5: excitation temperature derived
from $^{12}$CO. Col. 6: optical depth of the $^{13}$CO line. Col.
7 and 8: $^{13}$CO and H$_2$ column densities. Col. 9: total mass.
Col. 10: average H$_2$ density assuming a slab with $n=N({\rm
H}_2)/2/r$. Col. 11: equivalent radius determined by
r=(area/$\pi$)$^{0.5}$ and deconvolved with the beam size. $^a$
Taken from Arnal et al.~(\cite{arnal03}). $^b$ maximum line
temperature of $^{12}$CO 1$\to$0 $^c$ Taken from  Yamaguchi et
al.~(\cite{yamaguchi99}).\label{excitation}}
\begin{tabular}{lccccccccccccl}
(1) &  (2)  &  (3) &  (4) & (5) & (6) & (7) & (8) & (9) & (10) &  (11) \\
\noalign{\smallskip}\hline \noalign{\smallskip}
& T($^{13}$CO) & T($^{12}$CO)
& v & T$_{ex}$ & $\tau$ &  N($^{13}$CO) &  N(H$_2$) & Mass & n(H$_2$) & r \\
& [K] & [K] & [km s$^{-1}$] & [K] &  & [10$^{17}$ cm$^{-2}$]
&
[10$^{21}$ cm$^{-2}$] & [M$_\odot$] & [10$^3$ cm$^{-3}$] & [pc] \\
\noalign{\smallskip}\hline \noalign{\smallskip}
IRAS~16362-4845 &  & & & & & & & & & \\
\noalign{\smallskip}\hline \noalign{\smallskip}
Core         & 12.2 &  30.1 & --24.5 & 35.5 & 0.47 & 1.5 & 70 &  187 & 90.0 & 0.12 \\
Whole cloud  &  2.0 &  10.0 & --24.5 & 15.0 & 0.19 & 0.8 & 39 & 8000 &  3.9 & 1.62 \\
Cloud E$^a$  &   & 6.3$^b$  & --23.4 & 9.5  &      &     &    & 3900 &  4.4 & 3.56 \\
Cloud 78$^c$ & 6.7  &       & --22.5 &      & 0.52 &     & 25 & 4700 &  1.4 &   \\ 
\noalign{\smallskip}\hline \noalign{\smallskip}
NGC~6193 &  & & & & & & & & & \\
\noalign{\smallskip}\hline \noalign{\smallskip}
Whole cloud  &  2.0 &   8.6 & --16.5 & 13.6 & 0.22 & 0.4 & 18 & 660  &  4.5 &  0.66 \\
Cloud O$^a$  &   & 2.4$^b$  & --17.5 &  5.6 &      &     &    & 540  &  0.09 & 5.7 \\
\noalign{\smallskip}\hline
\end{tabular}
\end{table*}

\subsubsection{Channel Maps\label{channel}}

\noindent {\bf \underbar{Eastern  Cloud -- NGC 6193}} \\

  Figure~\ref{chan_ngc} shows a series of velocity channels of \co
(\cod) emission around NGC 6193. While the emission distribution
of \co (top) is highly fragmented and dispersed in all velocity
ranges, the \cod emission (bottom) focusses mainly on a single
region close to the pair of O stars in a rather narrow velocity
range ($\sim -$18 to $\sim -$15~km~s$^{-1}$).  However, there
is a clear morphological and kinematical coincidence between the
most prominent structures visible in both molecular species. As
can be seen in Fig. 4 in Arnal et al. (\cite{arnal03}), there are
molecular clouds further south of our mapped region at velocities
between $-$17.5 and $-$11.2~km~s$^{-1}$ which may constitute the
remains of an initially more extended molecular cloud, now mostly
disrupted by the NGC~6193 cluster.

\begin{figure}
\begin{center}
  \resizebox{8.5cm}{!}{\includegraphics[angle=-90]{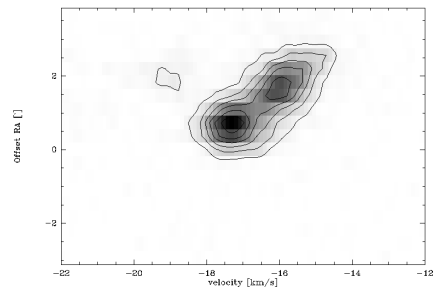}}
  \caption[]{
Position-velocity cut of \cod at constant declination (offset
1.$'$5) close to NGC~6193. Contours go from 1 to 7~K~km~s$^{-1}$
in steps of 1~K~km~s$^{-1}$. }
  \label{velopos}
\end{center}
\end{figure}

  The single --24~km~s$^{-1}$ velocity component visible in the
spectra of Fig.~\ref{moloverview} is prevalent across the whole
map and looks rather unrelated to the bulk emission of the cloud
which starts at --20~km~s$^{-1}$ and peaks at $-$17~km~s$^{-1}$.
There is a clear velocity shift in \cod emission (which is less
easily recognized in \co) in this region which becomes even more
obvious in a position-velocity plot shown in Fig.~\ref{velopos}.
This cut at constant declination (offset 1.$'$5) shows that the
velocity decreases from $\sim -$16~km~s$^{-1}$ to
$\sim -$18~km~s$^{-1}$ from east to west. The physical interpretation
of this gradient is not unambiguous. If caused by solid-body rotation
of the cloud, in which the cloud itself is continuous and turns
around the O stars whose wind/radiation created a hole around
them, the velocity gradient of 1.3~km~s$^{-1}$ across 1$'$ would
indicate a period of 1.7$\times$10$^7$~yr, which is in the same
order of magnitude like the one for the Rosette Molecular Cloud
(3.1$\times$10$^7$ yr, Blitz and Thaddeus~\cite{blitz80}) or the
Orion Nebula (Kutner et al.~\cite{kutner77}).

Expansion powered by the energetic output to the massive
stars in NGC~6193 provides an alternative explanation to the
observed structure and kinematics of this cloud, already suggested
by Phillips et al.~(\cite{phillips86}) and Yamaguchi et
al.~(\cite{yamaguchi99}). In this interpretation the arc-shaped
peak most clearly seen in the $-16$ and $-15.4$~km~s$^{-1}$ \co
channel maps of Fig.~\ref{chan_ngc} would be part of the rim of
an expanding shell containing HD~150135/150136 within its contour,
whereas the peak close to the positions of those stars and having
a somewhat more negative velocity ($\simeq -17$~km~s$^{-1}$) would
be part of the shell hemisphere located in front of the stars and
moving towards us. Difficulties with this interpretation may be
however indicated by a comparison of the momentum of the shell and
the the momentum injected by the winds of the O stars on their
surroundings during their lifetimes. The momentum of the cloud can
be estimated by taking the velocity difference between the
proposed frontal feature of the shell and the rim ($\sim
2$~km~s$^{-1}$) and multiplying it by our mass estimate of the
cloud (660~M$_\odot$; see Section~\ref{masses}). For the stellar
winds, we adopt typical mass loss rates (${\dot M} \sim
10^{-7}-10^{-8}$~M$_\odot$~yr$^{-1}$), terminal velocities
($v_\infty \sim 1000$~km~s$^{-1}$), and an age of $3.1 \times
10^6$~yr (V\'azquez \& Feinstein~\cite{vazquez92}). The momentum
contained in the expanding shell turns out to be greater than that
injected by the stars by about one order of magnitude. The actual
difference is probably greater, since the currently derived
660~M$_\odot$ seem to be only the remnants of a larger, now mostly
dispersed cloud. On the other hand it is also possible to invoke
particular density distributions that may improve the agreement in
our comparison, as would be the case if the proposed rim were
actually a high density ridge expanding at a lower speed than the
layer of gas moving towards us along the line of sight.
Unfortunately, the fragmentary nature of the cloud and the
tentative character of the interpretation of the structures
identified in it prevent us from deciding between both outlined
possibilities.

\noindent {\bf \underbar{Western Cloud -- IRAS~16362-4845}} \\
  The channel maps of the molecular cloud west of the bright rim
(Fig.~\ref{chan_rcw}) show two prominent features, firstly a sharp
gradient in \co and \cod line intensity at the interface region to
the rim, and secondly a pronounced peak of both, \co and \cod
emission close to the location of IRAS~16362-4845 (marked as a
triangle in the figure). Apart of these morphological elements the
western molecular cloud is not as strongly fragmented as the
eastern one. The main features of the molecular gas that we
identify near the position of the compact HII region are in
general coincident with those discussed by Urquhart et
al.~(\cite{urquhart04}). However, the greater extension of our
mapped area and the somewhat improved sensitivity in the
$^{12}{\rm CO} \ (J=2 \to 1)$ line allow us to carry our a more
comprehensive discussion on the structure and kinematics of the
surrounding regions as well.

The HII region NGC 6188 and the molecular cloud are in direct
contact in the velocity range $\sim$--25 to $\sim$--23~km~s$^{-1}$
where we see a sharp gradient in CO intensity at the interface.
Towards higher velocities the interface region moves westwards and
CO emission is only found close to the IRAS source. However, low
intensity emission -- best visible in the \co map -- remains
across the whole mapping area.  \co and \cod peak emission is
directly associated with IRAS~16362-4845 only in the velocity
range --26~km~s$^{-1}$ to --23.5~km~s$^{-1}$.  There, we find
$^{13}$CO column densities (N($^{13}$CO)=1.5$\times$10$^{17}$
cm$^{-2}$) at a moderate optical depth ($\tau$=0.8). Towards
higher velocities, the region of peak emission moves north-west
and is displaced by $\sim$40--60$''$ with regard to the IRAS
source. This core region is projected over a plateau of lower
intensity widespread \co emission. This lower column density
material is not visible in \cod where the region of peak emission
is more extended but also more clearly defined.  We do not see a
clear outflow signature in the \co channel maps (see
Sec.~\ref{overandspectra} for a discussion of the spectra) though
a bipolar structure with a blue component (--27.4 to
--26.3~km~s$^{-1}$) east of IRAS~16362-4845 and a more prominent
red component (--21.8 to --16.8~km~s$^{-1}$) on the western side
can possibly be identified. Higher angular resolution should
clarify to which extend this emission is due to YSO. So far we
attribute the redshifted high-velocity molecular gas to another
molecular cloud fragment which is eroded by the ionizing gas of
the compact HII region forming IRAS~16362-4845.

The morphological relationship between the molecular gas
mapped in this region and the structures seen in visible and
near-infrared images is obvious. In addition to the noted
correspondence between IRAS~16362-4845, its associated stellar
aggregate revealed by the infrared images
(Section~\ref{aggregate}), and the peak in molecular emission
intensity, the rim nebula corresponds to a steep gradient in
molecular gas density indicating that the molecular gas is
dynamically influenced by the presence of the ionization front
directly ahead of it. The associated shock front propagating into
the cloud is best visible for \cod at $v \simeq -24$~km~s$^{-1}$.
Further west, the CO emission distribution is more elongated and
looks like streaming away from the front at velocities
$\sim$--24.6 to $\sim$--21.3~km~s$^{-1}$.

\subsubsection{Physical properties of the molecular clouds\label{masses}}

  Table~\ref{excitation} gives an overview of the physical
conditions of the bulk emission of the molecular clouds derived
from our \co and \cod SEST observations and in comparison with
values from the literature.

  The excitation temperature was calculated by assuming an
optically thick $^{12}$CO line so that the radiative transfer
equation simplifies to $T_{ex}=11.06\times(\ln(11.06/(T(^{12}{\rm
CO})+0.187) + 1))^{-1}$ [K] with the line temperature T($^{12}$CO)
[K] determined with an Gaussian fit to the observed line.  The
$^{13}$CO line was assumed to be optically thin and the opacity is
then given by $\tau(^{13}{\rm CO})=-\ln(1-T(^{13}{\rm
CO})/(5.289/(\exp(5.289/T_{ex})-1)-0.868))$ with the line
temperature T($^{13}$CO) [K] equally determined with an Gaussian
fit. We assume LTE so that the excitation temperatures for
$^{12}$CO and $^{13}$CO are equal.  With the line integrated \cod
intensity W($^{13}$CO) [K~km~s$^{-1}$] over different areas (see
Table~1), the $^{13}$CO column density is given by $N(^{13}{\rm
CO}) [cm^{-2}] = f(T_{ex}) \times 10^{15} W_{^{13}{\rm CO}}$
(Frerking et al. \cite{frerking82}) with a value of $f(T_{ex})$ of
0.87, 1.06 and 1.87 at an excitation temperature of 10, 15 and 35
K.  The H$_2$ column densities are then calculated using the
$^{13}$CO column density with $N({\rm H_2})[cm^{-2}]=4.7\times10^5
N(^{13}{\rm CO})[cm^{-2}]$ (Dickman \cite{dickman78}).  The masses
were determined by $M[M_\odot]=6.6\times10^{-24} N({\rm H_2}) D^2
A$ with the distance of the cloud $D$ in parsec (1300 pc) and the
areal extent $A$ in square degrees. The average H$_2$ density is
evaluated by assuming a slab with a column of the length 2
$\times$ Radius.

  For the western cloud, we distinguish two different areas: the
cloud core correlated with  IRAS~16362-4845 is covered by
approximately one beam in \cod and the whole cloud is defined by
the 5$\sigma$ level of the observations (essentially all emission
visible in Fig.~\ref{moloverview}). The latter compares best with
the literature values.  The molecular core extends over
$\sim$0.24~pc which is approximately the same size as for the
embedded compact HII region plus halo (Sec.~\ref{morphology}). The
extinction given by the H$_2$ column density is very high ($A_V
=70^m$),  and when compared to the extinctions derived from the
stellar infrared colors (Section~\ref{aggregate}) it suggests that
the embedded aggregate lies roughly between the edge of the cloud
facing us and the center of the absorbing column of gas and dust
in its direction. The  volume density is high (9$\times$10$^4$
cm$^{-3}$) indicating the existence of a large reservoir of dense
molecular gas. The cloud core mass is $\sim$200 M$_\odot$, which
is similar to the stellar mass of the whole aggregate
(Section~\ref{aggregate}). The whole cloud has a mass of
$\sim$8000 M$_\odot$ which is approximately two times larger than
the masses derived by Arnal et al.~(\cite{arnal03}) while the area
is two times smaller. However, their observations are strongly
affected by beam dilution (8$'$.7 angular resolution) and
$^{12}$CO 1$\to$0 was used for the mass estimate which may explain
the differences in masses. Yamaguchi et al.~(\cite{yamaguchi99})
also use \cod as a mass tracer and obtain a larger total cloud
mass though at lower average densities.

  For the eastern cloud, we determined the cloud properties for
the region of strongest \cod emission visible in
Fig.~\ref{chan_ngc} in the velocity range --18 to --14 km
s$^{-1}$. While the optical depth ($\sim$0.2) is very similar in
comparison to the western cloud, the $^{13}$CO -- and accordingly
H$_2$ -- column densities are smaller. The gas is not as dense
($n \sim 4.5 \times 10^3$~cm$^{-3}$) as for IRAS~16362-4845 and
the total mass (660 M$_\odot$) is much lower than for the western
cloud. Also the extinction on the background, $A_V \simeq
18^m$, is much lower than in the western cloud. All that,
together with its more fragmentary morphology, points towards a
more evolved molecular cloud which constitutes now remnant
material.

\subsection{The kinematics of the ionized gas}

\begin{figure}
\begin{center}

\includegraphics[scale=0.4,angle=-90,bb=50 126 546 690,clip]{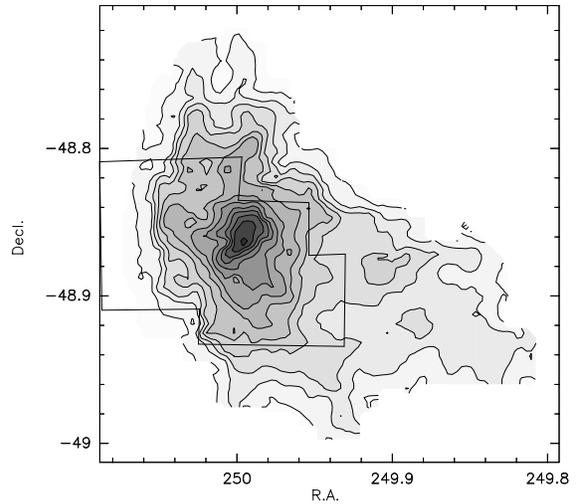}
  \caption[]{
Velocity integrated \co emission overlaid with the limit of the interferometric H$\alpha$ mapped area.}
  \label{pf1}
\end{center}
\end{figure}

\begin{figure}
\begin{center}
  \resizebox{8.5cm}{!}{\includegraphics[angle=-90]{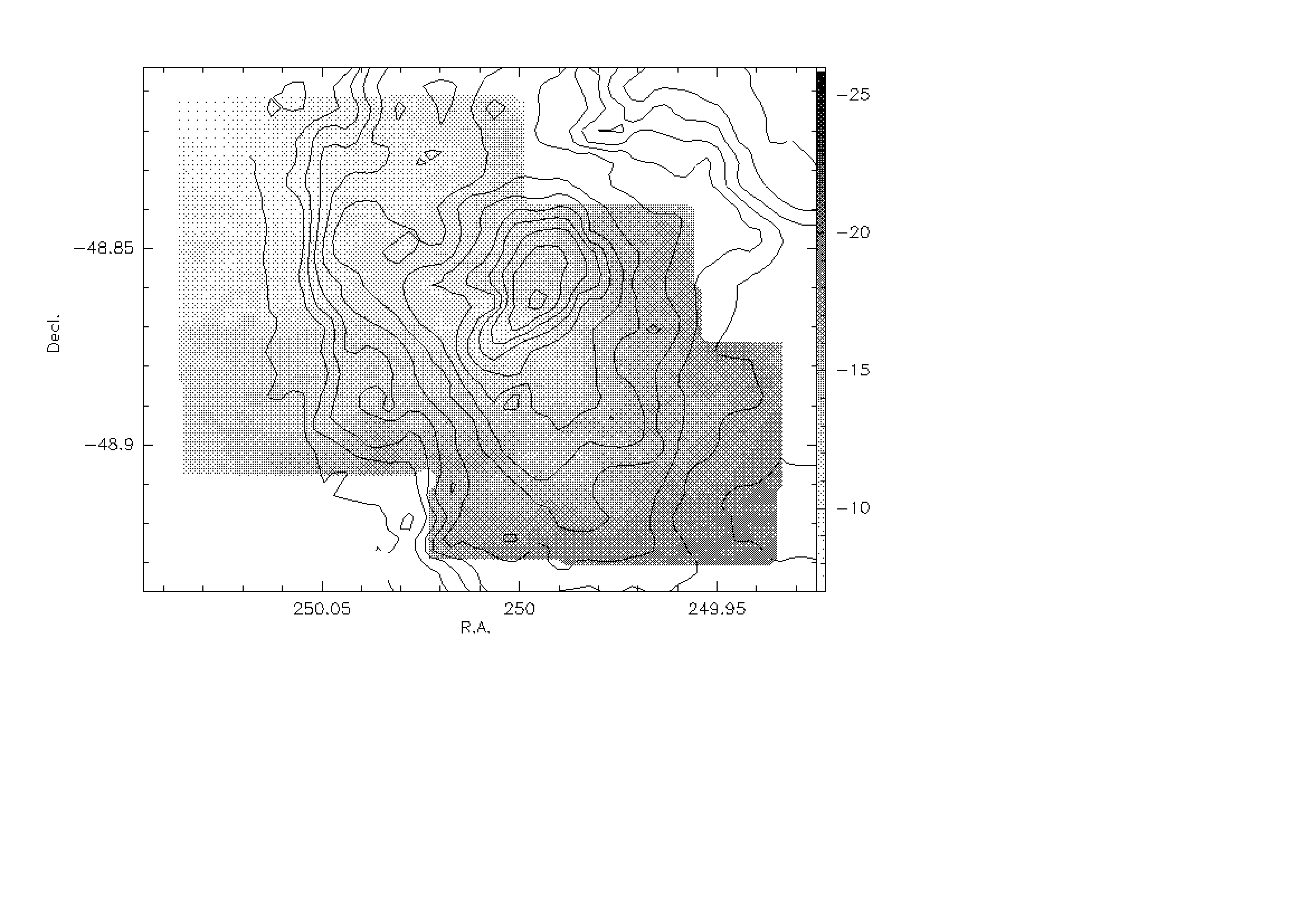}}
  \resizebox{8.5cm}{!}{\includegraphics[angle=-90]{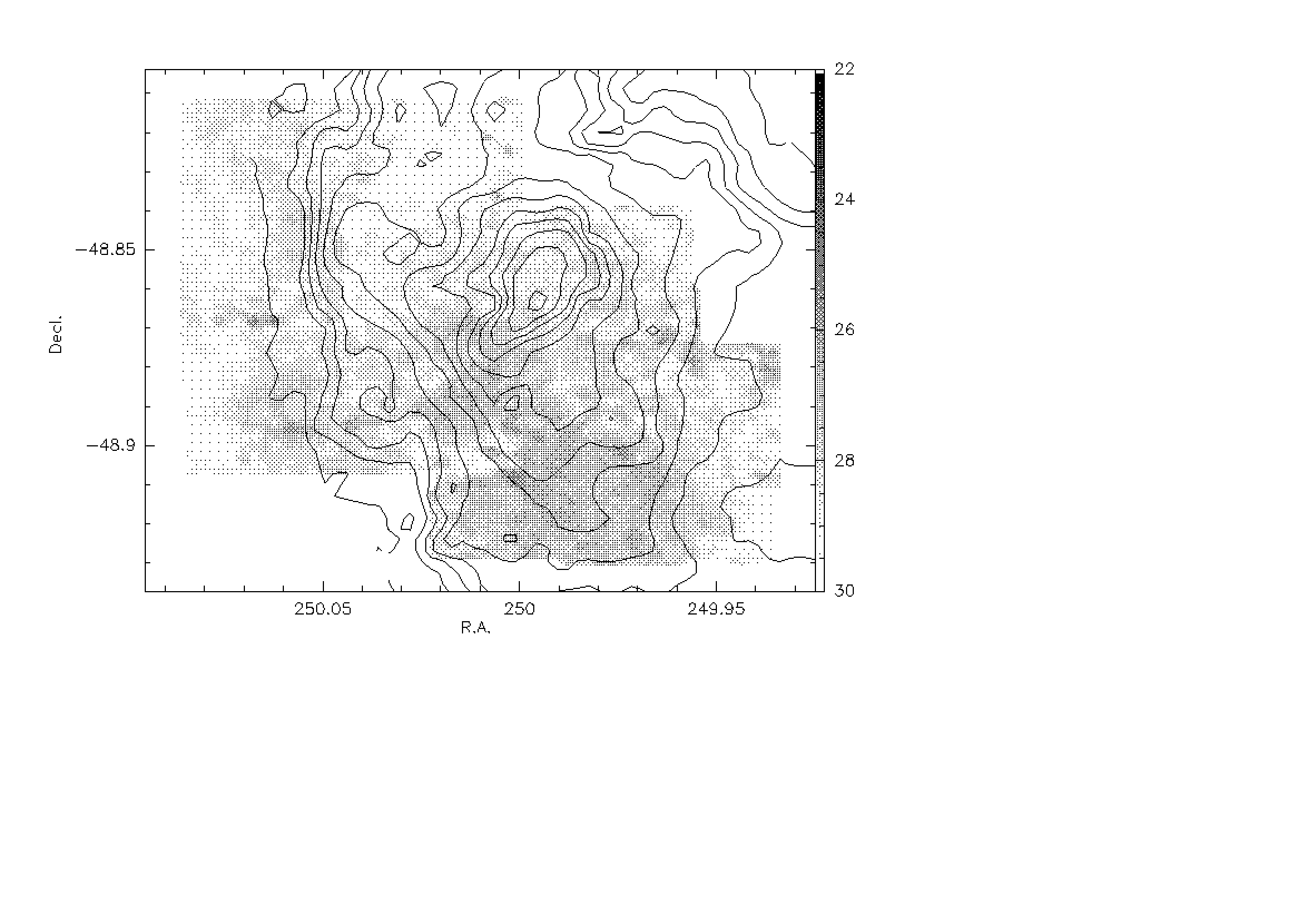}}
  \resizebox{8.5cm}{!}{\includegraphics[angle=-90]{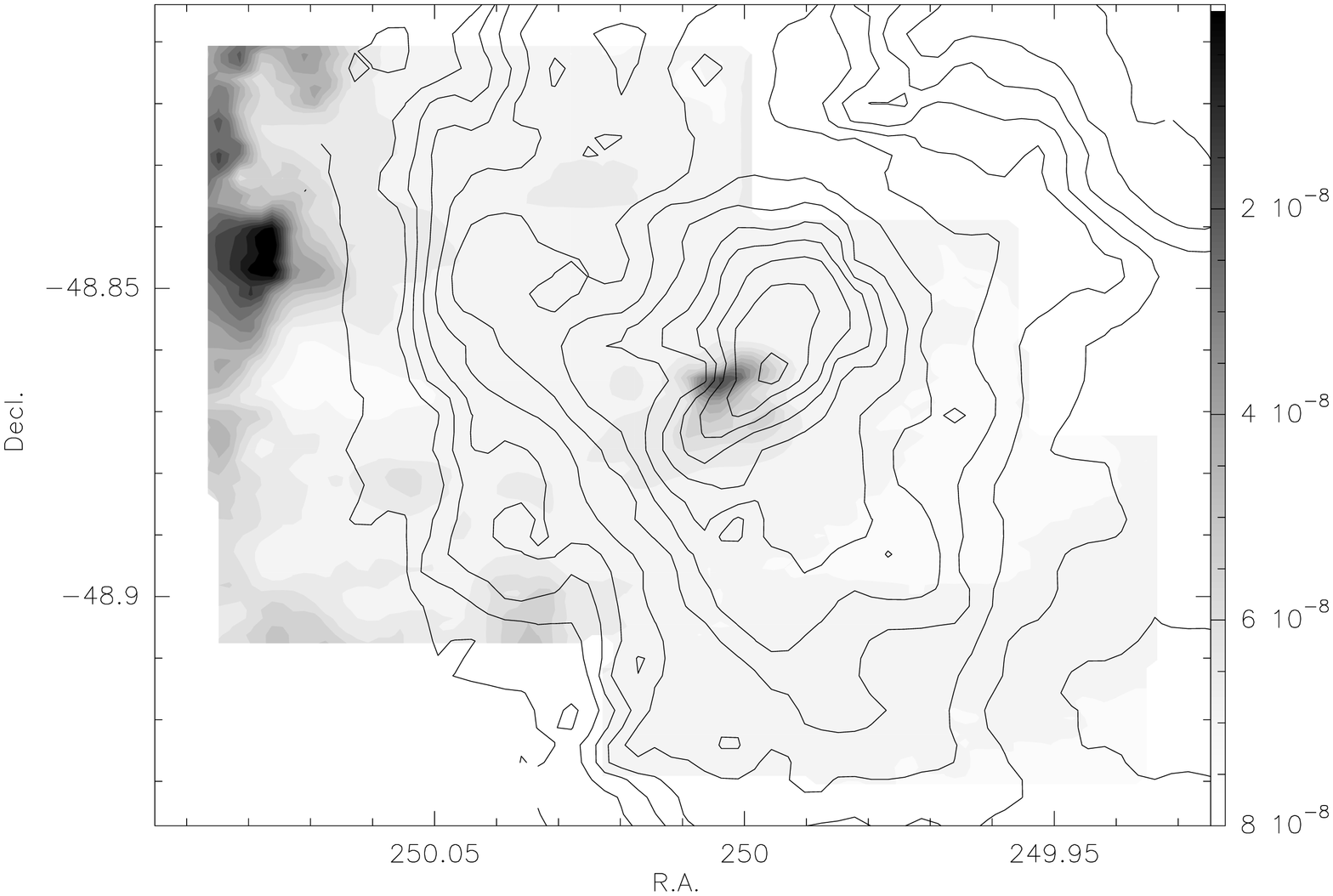}}
  \caption[]{
Central velocity (in km~s$^{-1}$), full width at half maximum (in km~s$^{-1}$) and intensity distribution
(in ergs/s/cm$^{-2}$/sr/km~s$^{-1}$). The overlaid isocontours correspond to the \co emission.}
  \label{pf2}
\end{center}
\end{figure}

The continuous H$\alpha$ information over the observed field
(Figure~\ref{pf1}) allows us to produce maps of the ionized gas
velocity, line width and intensity (Figure~\ref{pf2}). It was
always possible to fit the observed profiles by a single Gaussian.

The intensity map (Fig.~\ref{pf2}, bottom) clearly shows two major
features, the bright rim in the west and the compact HII region
in the map center. More diffuse emission is present in the rest
of the field. The compact HII region appears displaced from the
peak of the CO cloud. However we can note that it is located where CO
isocontours are distorded looking like a cavity from where the ionized
gas can expand.  This is a clear indication of the interaction between
the HII region and its parental molecular cloud.

\begin{figure}
\begin{center}

\includegraphics[scale=0.65,bb=15 440 417 830,clip]{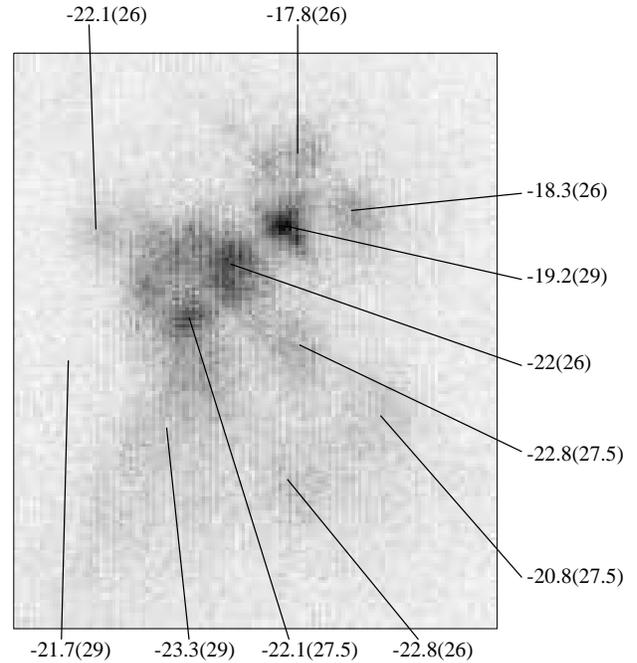}
  \caption[]{
Full resolution monochromatic H$\alpha$ image of the compact HII region.
For some strategic positions, the Vlsr, with the line widths in parentheses,
are indicated (in ~km~$\rm s^{-1}$). The image is centered on
$\alpha(2000) = 16^h 40^m 0\fs 3$, $\delta(2000) = -48^\circ 52' 06''$
and covers a field of $1'20 \times 0'97$, with North at the top and East
to the left.
}
  \label{pf3}
\end{center}
\end{figure}

As already mentioned section 3.1., the compact HII region is very
patchy and slightly elongated in South-East direction
(Figure~\ref{pf3}). Integrated over the whole compact HII region
the H$\alpha$ profile gives a systemic velocity of
$-21.7$~km~s$^{-1}$ (FWHM 27.5~km~s$^{-1}$). A detailed analysis
of the different patches highlights a small velocity variation in
the same South-East direction with velocities from --17.8 to
--23.3~km~s$^{-1}$. The brightest clump has a velocity of
--19~km~s$^{-1}$ (FWHM 29~km~s$^{-1}$). In the direction of the
opaque dust lane the H$\alpha$ velocity is also
--21.7~km~s$^{-1}$.  The bulge of CO emission emission at
--24.5~km~s$^{-1}$ can be associated to the extended emission of
the HII region while the H$\alpha$ patches can be counterparts of
the bumps around --20~km~s$^{-1}$ seen in the CO profile.  This
can be interpreted as dense clumps externally ionized and
photoevaporated. It is consistent with the conclusion inferred
from the infrared images where the core of the compact HII region
is deeply embedded in the molecular cloud and only the external,
less obscured parts are revealed in the visible.

The line width of the ionized gas (Fig.~\ref{pf2}, middle) shows a
small and marginally significant increase at the location of the
sharp edge of the molecular cloud (25~km~s$^{-1}$) while towards
the West and East the width is around 23~km~s$^{-1}$. This trend
is probably due to the presence of the ionization front. Elsewhere
in the mapped area and farther from the ioniation front, the mean
width is 27~km~s$^{-1}$.  The line width is the combination of two
factors, thermal gas motions and turbulence. A typical HII region
temperature of 10000 K gives a thermal FWHM of 21.4~km~s$^{-1}$.
If we interpret the width broadening by the turbulence this leads
to a velocity dispersion between 8.5 and 16.5~km~s$^{-1}$.
Systematic velocity dispersion trends and variations were already
reported in other HII regions (e.g. Godbout \cite{godbout97} and
references therein).

Finally, a clear large-scale velocity gradient (Fig.~\ref{pf2},
top) is visible. From velocities of --25~km~s$^{-1}$ in the North-East part
of the map, we go up to about --11~km~s$^{-1}$ in the South-West part
with a mean velocity --16.2$\pm$4.3~km~s$^{-1}$.  This gradient points
towards NGC~6193 confirming that it is the ionizing source. A streaming
motion of gas toward us can be excluded because such motion would imply
a velocity gradient with decreasing velocities. The velocities at the
ionization front up to the compact HII region are in good agreement
with the CO velocities. The H$\alpha$ emission at these velocities
certainly comes from the direct interaction between the ionizing flux
and the western molecular cloud.

The higher velocities correspond to more diffuse and fainter
H$\alpha$ emission except the structure around $\alpha (2000) =
16^h 40^m 08^s4$, $\delta(2000) = -48^\circ 54' 18''$ (--14 km
s$^{-1}$) which is quite intense. The Western cloud does not show
similar velocities but such velocities are noted for the Eastern
cloud ($\sim$ --15~km~s$^{-1}$). In parallel, Arnal et al.
(\cite{arnal03}) detected molecular clouds further south at
velocities between $-$17.5 and $-$11.2~km~s$^{-1}$ which are
supposed to be remains of an initially more extended molecular
cloud, now mostly disrupted by the NGC~6193 cluster.  Hence the
H$\alpha$ emission at velocities between --16 and --11~km~s$^{-1}$
is probably the ionized counterpart of this initial parental
cloud.  However due to the limited extension of the H$\alpha$ data
it is not possible to arrive to definite conclusions.

\subsection{The ionizing aggregate of IRAS~16362-4845\label{aggregate}}

  The stellar component of IRAS~16362-4845 is revealed by the
$JHK_S$ images, where a tight aggregate of stars emerges at the
center of the compact HII region. As noted in
Section~\ref{masses} the aggregate approximately coincides in
position with the peak in column density of molecular gas, which
causes an extinction on the background of $A_V = 70^m$ (or $A_K =
8^m$), thus ruling out that it may be actually composed of
background stars seen through the cloud.

\begin{figure}
  \resizebox{8.5cm}{!}{\includegraphics{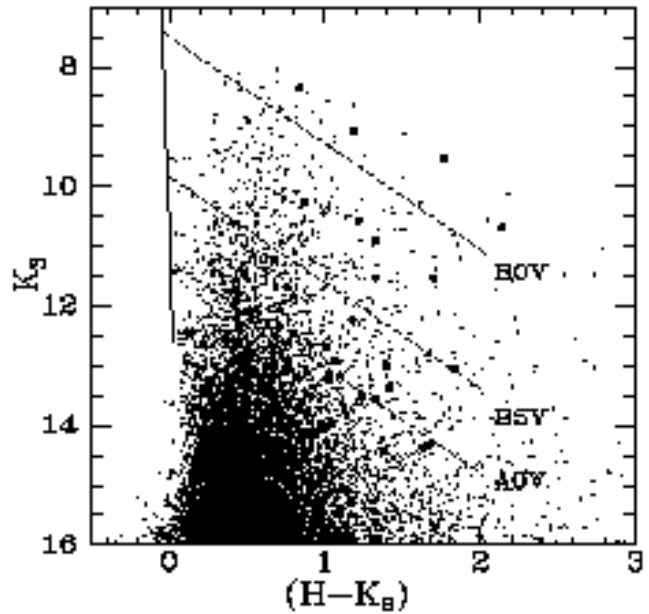}}
  \caption[]{Color-magnitude diagram of all the objects detected
in the infrared mosaic images. The filled circles indicate the
stars lying projected on the central $40''\times 40''$ of the
nebula (see also Table~\ref{clustermembers}). The dashed lines
mark the expected positions of stars of different spectral types
(whose unreddened colors and magnitudes are given by the
nearly-vertical solid line at the distance of Ara~OB1) reddened by
varying amounts. The range of $(H-K_S)$ colors suggest a wide
range of extinctions of the stars embedded at different depths in
the nebula, ranging from $A_V \simeq 8$ to $A_V \simeq 35$.}
  \label{colmag}
\end{figure}

\begin{figure}
  \resizebox{8.5cm}{!}{\includegraphics{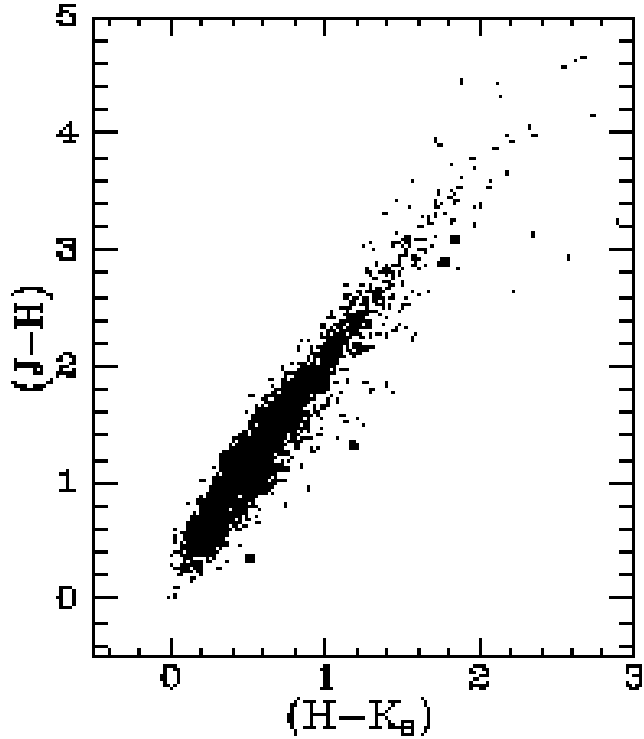}}
  \caption[]{Color-color diagram of all the objects detected
in the infrared mosaic images. To avoid degrading the quality of
the diagram with faint objects whose colors are uncertain we have
limited the sample plotted here to objects brighter than $K_S =
14.5$. The filled circles indicate the stars lying projected on
the central $40''\times 40''$ of the nebula (see also
Table~\ref{clustermembers}).}
  \label{colcol}
\end{figure}

  The color-magnitude and color-color diagrams are useful tools in
examining the membership, rough spectral types, and possible
existence of infrared excess of the stars that appear projected on
the nebula. The color-color diagram of the aggregate at the core
of the HII region and its surroundings has been discussed by
Urquhart et al.~(\cite{urquhart04}), who have identified several
infrared-excess objects based on 2MASS data. Our SOFI data reach
over two magnitudes deeper, have a better linear resolution by a
factor of at least two, and oversample the PSF area. Therefore,
they allow us to obtain a much more detailed view of the stellar
contents of the aggregate. Both the color-magnitude and
color-color diagrams are presented in Figures~\ref{colmag} and
\ref{colcol} for the entire catalog of objects detected in our
near infrared mosaic of the region, with the objects projected on
the central $40'' \times 40''$ ($0.3$~pc~$\times$~$0.3$~pc) of the
infrared nebula marked with full circles. The position of the
members of the aggregate in the color-magnitude diagram shows that
the brightest stars seem to have late O types, and are accompanied
by several early B-type stars. The tightest concentration occurs
in a small cluster precisely at the position of IRAS~16362-4845,
and occupies a projected area approximately $8''$ across
(0.05~pc). Judging from their $(H-K_S)$ colors they are reddened
by widely varying amounts ranging from $A_V \simeq 8$ to $A_V
\simeq 29$ or perhaps $A_V = 35$ (depending on whether the reddest
star is a true member of the cluster or rather a background star;
see discussion below), using the Rieke \&
Lebofsky~(\cite{rieke85}) extinction law and assuming essentially
zero $(H-K_S)$ intrinsic colors. The compactness of the aggregate
and its stellar contents are reminiscent of a Trapezium-like
cluster. Other stars having $(H-K_S)$ colors in that same range
are visible at some distance from the central cluster but still
within the confines of the compact HII region.

\begin{figure}
  \resizebox{8.5cm}{!}{\includegraphics{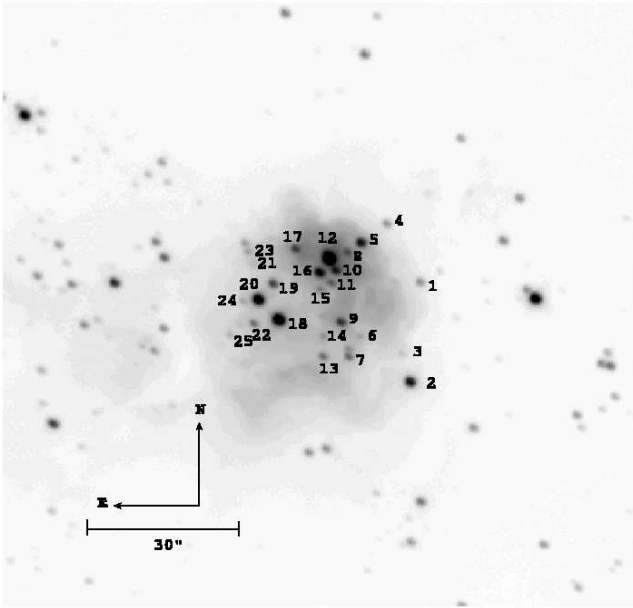}}
  \caption[]{$K_S$-band image of the IRAS~16362-4845 aggregate,
indicating the numbers of the objects whose positions and
magnitudes are given in Table~\ref{clustermembers}.}
  \label{clustermap}
\end{figure}

\begin{table*} \caption{Positions and photometry of stars projected
on the HII region\label{clustermembers}}
\begin{tabular}{cccccccl}
\noalign{\smallskip}\hline \noalign{\smallskip}

Number & $\alpha(2000)$ & $\delta(2000)$ & $K_S$ & $(J-H)$ & $(H-K_S)$ & ${A_V}^1$ & Notes \\
\noalign{\smallskip}\hline \noalign{\smallskip}

  1 & 16:39:58.4 & -48:51:45 & 12.931 &  2.137 &  1.080 & 17.3 & B5-A0, no excess      \\
  2 & 16:39:58.6 & -48:52:05 & 10.568 &  2.163 &  1.217 &      & B0-B5, slight excess      \\
  3 & 16:39:58.8 & -48:51:59 & 14.363 &   -    &  1.637 &      & later than A0      \\
  4 & 16:39:59.0 & -48:51:34 & 13.044 &  3.078 &  1.831 &      & B5 or somewhat earlier, slight excess      \\
  5 & 16:39:59.6 & -48:51:37 & 10.679 &   -    &  2.144 &      & late O, or perhaps background       \\
  6 & 16:39:59.7 & -48:51:56 & 14.030 &   -    &    -   &      & most probably B5-A0    \\
  7 & 16:39:59.8 & -48:52:00 & 12.236 &  2.370 &  1.179 & 18.9 & B5-A0, no excess      \\
  8 & 16:39:59.8 & -48:51:39 & 12.092 &  0.345 &  0.508 &      & possible excess; later than A0\\
  9 & 16:40:00.0 & -48:51:53 & 11.230 &  1.136 &  0.673 & 10.9 & B5-A0, no excess      \\
 10 & 16:40:00.1 & -48:51:42 & 10.907 &   -    &  1.324 &      & B0-B5      \\
 11 & 16:40:00.1 & -48:51:45 & 12.701 &   -    &  1.008 &      & B5-A0      \\
 12 & 16:40:00.2 & -48:51:40 &  8.356 &  1.674 &  0.842 & 14.0 & late O, no excess      \\
 13 & 16:40:00.3 & -48:51:59 & 12.449 &  0.520 &  0.124 &      & probably foreground; in any case, later than A0 \\
 14 & 16:40:00.3 & -48:51:56 & 13.528 &   -    &  1.247 &      & A0 \\
 15 & 16:40:00.4 & -48:51:46 & 13.001 &   -    &  1.400 &      & B5-A0    \\
 16 & 16:40:00.4 & -48:51:43 & 10.276 &  1.921 &  0.872 & 14.0 & B0-B5, no excess, CO and Br$\gamma$ in the spectrum \\
 17 & 16:40:00.9 & -48:51:39 & 11.537 &   -    &  1.702 &      & B0-B5     \\
 18 & 16:40:01.2 & -48:51:52 &  9.074 &  1.320 &  1.181 &      & probably earlier than B0, but strong excess \\
 19 & 16:40:01.3 & -48:51:45 & 11.525 &  2.617 &  1.329 & 21.3 & B0-B5, no excess \\
 20 & 16:40:01.6 & -48:51:48 &  9.529 &  2.882 &  1.765 & 28.7 & late O, perhaps slight excess; background? \\
 21 & 16:40:01.7 & -48:51:41 & 14.318 &   -    &  1.682 &      & somewhat later than A0  \\
 22 & 16:40:01.7 & -48:51:53 & 12.451 &  1.772 &  0.793 & 12.8 & B5-A0, no excess   \\
 23 & 16:40:01.8 & -48:51:39 & 13.379 &   -    &  1.415 &      & B5-A0 \\
 24 & 16:40:01.9 & -48:51:48 & 13.567 &   -    &  1.332 &      & B5-A0 \\
 25 & 16:40:02.2 & -48:51:55 & 14.436 &   -    &  1.383 &      & later than A0 \\

\noalign{\smallskip}\hline
\end{tabular}
$^1$: Extinction estimate using Rieke \& Lebofsky~(\cite{rieke85})
extinction law and the $(H-K_S)$ color index. The value is given
only for stars having $JHK_S$ measurements whose positions in the
color-color diagram indicates no infrared excess. Intrinsic colors
$(H-K_S)_0 = -0.04$ and $(H-K_S)_0 = -0.01$ are assumed for O
stars and stars in the B5-A0 range, respectively.
\end{table*}

\subsubsection{Individual members\label{ind_members}}

  The census of stars in the central $0.3 \times 0.3$~pc$^2$ of the
HII region is given in Table~\ref{clustermembers}, and plotted in
Figure~\ref{clustermap}. Some of these stars may be either
foreground or background, and thus unrelated to RCW~108. In
particular we consider Star~13, with blue $JHK_S$ colors, as a
very likely foreground source. In principle we also would be
inclined to consider as a foreground source Star~8, the one that
appears brightly in visible images nearly coincident with the
position of IRAS~16362-4845. However, its $(H-K_S)$ is much redder
than expected given the $(J-H)$ color, hinting at the existence of
a near infrared excess indicative of youth and thus possible
aggregate membership. Its position in the $(H-K_S, K_S$) diagram
indicates that it is unlikely to have a spectral type earlier than
A0, as confirmed by its visible spectrum, dominated by Balmer
lines and by the CaII H and K lines. Still near the cluster core,
Star~5 has $H$ and $K_S$ magnitudes consistent with a late O type.
We note however that its $(H-K_S)$ color implies an extinction
$A_V=35$ in its direction if entirely due to foreground reddening,
which would be one of the highest found among the objects in the
whole field, although only half that inferred from molecular-line
observations. Unfortunately, the star is too obscured for its $J$
magnitude to be measurable and no infrared spectrum is available
for us to decide on the actual membership of this star.

\begin{figure}
  \resizebox{8.5cm}{!}{\includegraphics{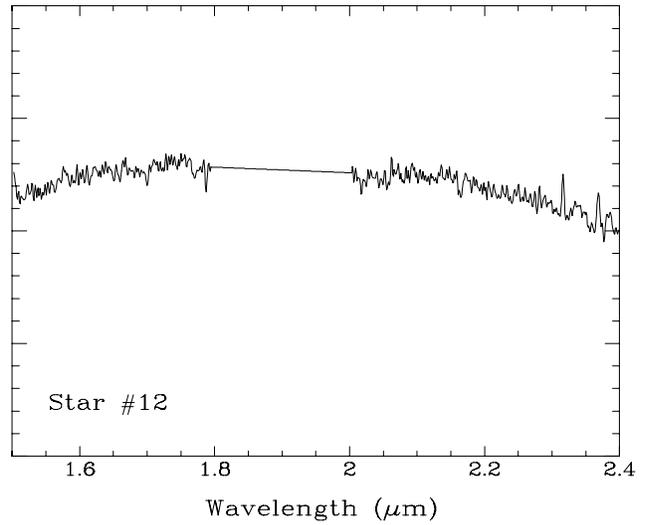}}
  \caption[]{Infrared spectrum of Star~12
(Table~\ref{clustermembers}), the most likely responsible for the
ionization of the IRAS~16362-4845 nebula. As discussed in the
text, the Br$\gamma$ absorption (2.166~$\mu$m) seems to be real.
This may also be the case for the two faint emission lines at
2.314 and 2.393~$\mu$m. The interval between 1.8~$\mu$m and
2.02~$\mu$m has been removed due to the strong telluric
absorption.}
  \label{star12_irspec}
\end{figure}

  Star~12 seems to play the main role in ionizing the
IRAS~16362-4845 nebula, judging from its central position at the
core of the nebula and its magnitude. Its photometry is in
agreement with the late-O spectral type that is expected from the
excitation characteristics of the nebula
(Section~\ref{spec_nebulae}). The fact that IRAS~16362-4845
contains a tight cluster rather than a single star explains the
discrepancy between the Lyman continuum flux and the luminosity
derived from far-infrared data pointed out by Straw et
al.~(\cite{straw87}). Star~12 is detected in visible-red images of
the region as a very red object next to Star~8, which dominates at
visible wavelengths. The visible spectrum of Star~12 shows only a
steep rise towards the red but no features, and is severely
contaminated by the nebula at the position of H$\alpha$ and other
strong nebular lines. The 1.5-2.4~$\mu$m spectrum, shown in
Figure~\ref{star12_irspec}, is also mostly featureless as expected
from a late O star. The hint of Br$\gamma$ weakly in absorption
must be taken with caution, as it may be an artifact due to
small-scale structure in brightness of the nebula affecting the
subtraction of the nebular lines from the spectrum of the star.
However we note that the subtraction process removes well the HeI
(2.058~$\mu$m) line, whose intensity in the nebula is similar to
that of Br$\gamma$ (Section~\ref{spec_nebulae}), leading us to
favor the interpretation of the Br$\gamma$ absorption as a real
feature in the photosphere of the star. Moreover, its strength is
consistent with that expected for a late-O spectral type (Hanson
et al.~\cite{hanson96}).

  It is interesting to note the possible existence of two
faint emission lines in the spectrum of Star~12 at $\lambda =
2.314~\mu$m and $\lambda = 2.393~\mu$m. Despite the increased sky
thermal background on the long-wavelength end of the $K$ band the
quality of the spectrum of Star~12 seems to be sufficiently good
at the position of these two lines, especially at 2.314~$\mu$m. We
have attempted several slightly differing reductions of the
spectra, by choosing different background subtraction apertures
and by excluding each one of the four spectra obtained at a time,
in order to ascertain the reality of the features, and in all
cases we have recovered them. The position of the 2.314~$\mu$m
feature matches well that of a feature seen in evolved massive
stars, such as WRA~751 (Morris et al.~\cite{morris96}), where it
is nevertheless accompanied by other, much stronger emission
lines. On the other hand, the good subtraction of all other
nebular lines suggests an origin in the photosphere or the wind of
Star~12. Therefore it is unfortunate that we cannot offer an
explanation for these features, and mention them here as a
possible puzzle should better quality observations provide a
definitive confirmation of their existence.

\begin{figure}
  \resizebox{8.5cm}{!}{\includegraphics{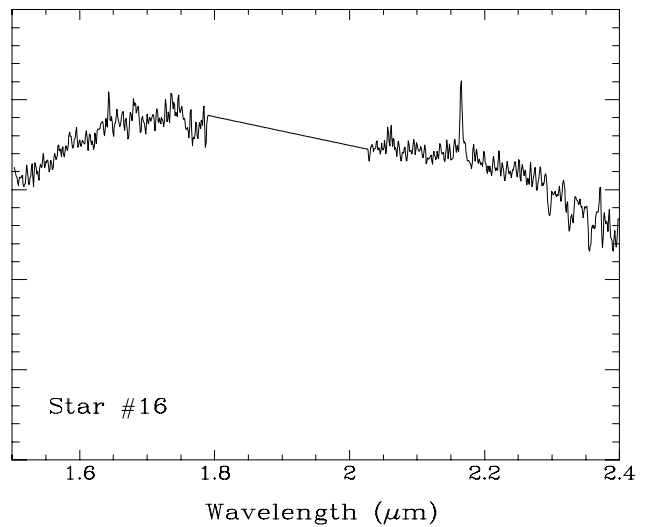}}
  \caption[]{Infrared spectrum of Star~16, showing both its
prominent Br$\gamma$ emission and the CO bands longwards of
2.29~$\mu$m. The interval between 1.8~$\mu$m and 2.02~$\mu$m has
been removed due to the strong telluric absorption. }
  \label{star16_irspec}
\end{figure}

  Star~16 is almost 2~mag fainter than nearby Star~12 but
has similar colors, suggesting that both are embedded at similar
depths in the cloud. Our infrared spectra had the slit oriented so
as to contain both stars, and the resulting spectrum of Star~16 is
shown in Figure~\ref{star16_irspec}. Examination of the frames
containing the spectral trace convincingly shows that the
prominent Br$\gamma$ emission in the spectrum of Star~16 is
related to it, and is not due to a poor subtraction of the nebular
lines of the surrounding HII region. The spectrum also clearly
shows the CO absorption bands starting at 2.29~$\mu$m typical of
cool atmospheres (e.g. Kleinmann \& Hall~\cite{kleinmann86}) and
suggest that Star~16 might actually be a background cool star.
However, the simultaneous appearance of both Br$\gamma$ emission
and CO absorption is frequent among young stellar objects
surrounded by significant amounts of circumstellar gas and dust
(Greene \& Lada~\cite{greene96}), which produce a much closer
match to the spectrum of Star~16. Our data also hint at color
variability of this source, as may be seen by comparing the
spectra in Figures~\ref{star12_irspec} and \ref{star16_irspec}
with the photometry in Table~\ref{clustermembers}. The imaging
observations yield a redder $(H-K_S)$ color for Star~16 than for
Star~12, while the slopes of their $HK$-band spectra, obtained
three years later, show that Star~16 was slightly bluer than
Star~12 in 2003.

  The photometry of Stars~18 and 20 also suggests a spectral type
earlier than B. Star~18 may indeed be a O-type star, but the
strong infrared excess derived from its $JHK_S$ colors, already
noted by Straw et al.~(\cite{straw87}), indicates that a
substantial fraction of the flux at $K_S$ may come from a disk,
and that the central star may actually be of later type. In either
case, the infrared excess can be taken as an indicator of its
youth and thus of membership in the aggregate. The situation of
Star~20 is less clear, as it displays a slight infrared excess at
most, probably insufficient to explain its brightness at $K_S$ as
dominated by circumstellar emission. Although it seems to be the
intrinsically brightest star of the IRAS~16362-4845 aggregate, no
obvious effects on the nebula are seen in its immediate
surroundings, thus leaving open the possibility that it could be a
background giant instead. We note however that the numerous giants
seen in the whole field imaged in the infrared trace a
well-defined reddening vector, from which Star~20 is detached by
nearly 0.3~mag. The sense of the departure from the {\it locus}
traced by giants is the same as observed in large-amplitude
variables (e.g. Glass et al.~\cite{glass95}). This possibility was
considered also by Straw et al.~(\cite{straw87}), whose
low-resolution spectrum of Star~20 ($=$~IRS~19 in their list) does
show indeed CO absorption at 2.3~$\mu$m. While considering it
unlikely on absolute magnitude grounds that Star~20 could be a
background giant or supergiant, they favored instead actual
membership in the aggregate and considered possibilities such a
T~Tauri star or even a FU~Ori object. Further infrared
spectroscopy of this star should clarify its nature.

\subsubsection{Stellar mass and density of the aggregate \label{star_mass}}

  It is possible to obtain a rough estimate of the stellar mass
of the IRAS~16362-4845 aggregate by assuming that our images are
sufficiently deep to record all the members with masses above a
certain threshold, and then extrapolating the mass function to
lower masses to account for the fainter, undetected members. Since
the distribution of colors and magnitudes of stars in the
aggregate plotted in Figure~\ref{colmag} suggests that there are
practically no stars earlier than A0 that may have been missed by
our survey (other than those that may be members of unresolved
binary pairs), we will use the mass corresponding to that spectral
type to derive the scaling factor of the mass function, for which
we assume a log-normal Miller-Scalo form (Miller \&
Scalo~\cite{miller79}). Table~\ref{clustermembers} indicates that
there appear to be approximately 19 stars with spectral types A0
or earlier, among which we include Star~20 as a member but exclude
Star~5 as a possible background star. The only star discarded as
foreground, Star~13, should be later than A0 if it is actually an
unobscured member of the aggregate and thus does not enter our
census. Similarly, the lightly reddened Star 8 is excluded from
the counting also due to its confirmed later spectral type. We
adopt $M = 2.9$~M$_\odot$ for a A0 main sequence star from the
compilation by Drilling \& Landolt (\cite{drilling00}), noting
that the spectral type-mass relationship may not be extended
to later spectral types given that stars with lower masses take at
least a few million years (5~Myr for a 2.5~M$_\odot$ star;
D'Antona \& Mazitelli~\cite{dantona94}) before reaching the main
sequence.

  The mass of the aggregate that we estimate in this way is
$\sim 210$~M$_\odot$, with a considerable uncertainty due to a
number of reasons. Our assessment of which stars are members and
which ones are foreground may be incorrect in some cases, thus
altering the true census of aggregate members. Moreover, as we
noted above, we have not made any corrections for unresolved
binarity, which is known to affect a large fraction of massive
stars (e.g. Garmany et al.~\cite{garmany80}). Also, the scaling
factor of the initial mass function is derived from its upper end,
which is affected by small-number statistics, and some stars may
appear above the A0 limiting line in Figure~\ref{colmag} because
of the existence of $K_S$-band excess. The evolutionary status of
the aggregate may also affect our mass estimate in a manner
similar to that described by Herbig \& Terndrup~(\cite{herbig86})
for the Trapezium cluster in Orion: the high brightness of some of
the stars that we detect may be due to their pre-main sequence
status rather than to their mass, resulting in luminosities higher
than those of main-sequence stars of the same masses, leading to
an overestimate of the number of massive members. This effect
should not affect the most massive, O-type stars like Star~12,
whose high temperature is confirmed by the spectrum of the HII
region, as such stars have reached already the main sequence by
the time that they become visible (Palla \&
Stahler~\cite{palla90}, Beech \& Mitalas~\cite{beech94}), but may
result in a systematic overestimate of the mass of other cluster
members and therefore of the cluster as a whole. Finally, we note
that the aggregate seems to contain numerous stars of spectral
type A0 and earlier with $(H-K_S) < 1.8$, but very few are found
below this line, and none with $H-K_S$ bluer than $(H-K_S) \simeq
1.2$ other than the likely foreground Star~13. No bias in our
observations can explain the lack of objects in the aggregate in
this region of the color-magnitude diagram (which is well crowded
for other parts of the field), and we can only explain it as a
real absence of moderately reddened stars with luminosities below
that corresponding to a main sequence A star. The apparent absence
of stars with masses below 2.5~M$_\odot$ was already noted by
Straw et al.~(\cite{straw87}), who suggested bimodality in the
mass function as a possible explanation. An alternative intriguing
possibility to explain these observations might be that at the
early age of the aggregate only the most massive stars, which
complete their evolutionary tracks towards the main sequence much
faster than intermediate-mass and solar-type stars, may be
sufficiently evolved and emerged from their circumstellar
envelopes to populate the region of the color-magnitude diagram
corresponding to moderate extinctions. Our observations do not
allow us to test this hypothesis, but future observations at
longer wavelengths and high spatial resolution may be able to do
it. Both the possible overpopulation of the region above the A0
main-sequence line in Figure~\ref{colmag} due to the early
evolutionary stage of the cluster, and the apparent lack of low
mass stars, may indicate an actual mass of the cluster below the
210~M$_\odot$ derived above. In the extreme case that no
stars less massive than 2.9~M$_\odot$ existed at all in the
cluster, and assuming that the likely members listed in
Table~\ref{clustermembers} have luminosities near the
main-sequence ones, the mass of the observed cluster population
would amount to $\sim 110$~M$_\odot$.

  It is interesting to compare the mass, contents, and extent
of the IRAS~16362-4845 aggregate with that of the best studied
young massive cluster, the Trapezium. The mass of the Trapezium
can be estimated at $\sim 130$~M$_\odot$ from the stellar mass
density and the approximate radius given by Herbig \&
Terndrup~\cite{herbig86}, which are respectively
3000~M$_\odot$~pc$^{-3}$ and 0.22~pc. Higher stellar densities are
obtained when considering only the central region of the cluster
(McCaughrean \& Stauffer~\cite{mccaughrean94}). The Trapezium
census of the most massive members includes 3 O-type members among
the components of $\theta^1$ and $\theta^2$~Ori (Warren \&
Hesser~\cite{warren77}), some of which are binaries having also
high-mass companions (Weigelt et al.~\cite{weigelt99}, Petr et
al.~\cite{petr98}). The mass that we obtain for the
IRAS~16362-4845 cluster appears to be somewhat higher, within the
caveats described above, and the number of O stars is similar,
again within the small-number statistics. It should be
pointed out that the Trapezium does contain a substantial
population of less massive members (McCaughrean \&
Stauffer~\cite{mccaughrean94}) such as solar-type stars, low-mass
stars, and substellar objects, in apparent contrast to the hints
of a significant lack of stars later than A0 that we have stressed
above. The luminosity function of the Trapezium cluster does not
suggest any marked deficiencies over any mass interval (Muench et
al.~\cite{muench02}), although noticeable effects of the most
massive stars on the mass function of the cluster have been
recently suggested by Robberto et al.~\cite{robberto04}). On the
other hand, independently of the actual stellar content the
IRAS~16362-4845 cluster is more compact. We estimate a radius of
0.11~pc, only half of the Trapezium cluster.  If the
extrapolation of the mass function to lower masses described
earlier is approximately correct, the mass density is
consequently higher than in the Trapezium by more than one order
of magnitude, nearly $4 \times 10^4$~M$_\odot$~pc$^{-3}$, which
translates into $n_{\rm H_2} = 6.4 \times 10^5$~cm$^{-3}$ assuming
that the mass forming the cluster was initially in the form of a
single molecular core. We note finally that the Trapezium is
probably nothing more than the nucleus of a much more extended
aggregate containing over 2000~M$_\odot$, the Orion Nebula Cluster
(Hillenbrand \& Hartmann~\cite{hillenbrand98}), for which no
counterpart exists in RCW~108 (See Section~\ref{efficiency}).

\subsection{Spectroscopy of the HII region\label{spec_nebulae}}

  The long-slit spectroscopy of selected stars in the visible and the
infrared also provides a cross section spectrum of the HII region
containing useful information on the physical conditions of its
different components and on the ionizing stars, complementing that
obtained with the observations of the stellar aggregate discussed
in the previous Section.

  The slit position in the visible spectrum contains stars 8 and 12,
and the slit length reaches well up to the edge of the
western molecular cloud, including a portion of the bright rim
nebulosity appearing at the erosion interface. It thus samples
well the compact component of the HII region, as well as the rim
nebula and the tenuous ionized gas foreground to the molecular
cloud that pervades the whole region. Moreover, when inspecting
the spectroscopic frames we have appreciated the existence of a
concentrated knot of emission in the visible appearing where the
slit position runs closest to Star~17. This knot, whose outskirts
are included in the slit, can be seen in the $V$-band image
(Figure~\ref{from_vis_to_ir}) and more clearly in the $R$-band
image. The compact, almost point-like feature in the visible
images is not coincident with Star~17, which is seen in the
infrared images only and is actually located 2''0 to the
northeast; both can be simultaneously seen only in the $J$-band
image of Figure~\ref{from_vis_to_ir}.

  Our estimates of the density of the ionized gas are made on the
basis of [SII]$(\lambda 6716) / (\lambda 6731)$ line ratio
(Osterbrock~\cite{osterbrock89}), and the estimates of the
temperature of the ionizing radiation are based on the single-star
photoionization models of Stasi\'nska \&
Schaerer~(\cite{stasinska97}) for a gas of solar metallicity and
for the density estimated from the [SII] line ratios. The
translation of an ionizing radiation temperature into a spectral
type of the ionizing star is made using the ZAMS models of
Schaerer \& de Koter (\cite{schaerer97}). We have estimated the
extinctions using either the intrinsic H$\beta$/H$\alpha$ line
ratio (foreground nebulosity and rim nebula) or the intrinsic
Paschen~6/H$\alpha$ line ratio (compact nebula and emission knot
near Star~17), taken from Osterbrock~(\cite{osterbrock89}) for
case B recombination. The extinction curve used is that of
Cardelli~et al.~(\cite{cardelli89}). We have also considered the
empirical extinction curve derived by Bautista et
al.~(\cite{bautista95}) for the Orion nebula, under the
expectation that it might better represent the extinction in the
direction of the compact nebulosity of IRAS~16362-4845 as well.
However, in the cases where the extinction can be simultaneously
measured from the Balmer and the Paschen line ratios we find
highly discrepant values between both when using Bautista et
al.'s~(\cite{bautista95}) curve, and a much better agreement when
using Cardelli et al.'s~(\cite{cardelli89}) curve, hence our
preference for the latter. It must be recalled anyhow that the
observed spectrum at any position is the integration along the
line of sight of the spectra produced over a range of depths,
extinctions, and conditions, and that the concept of typical
extinction towards any ionized component of the region is thus
necessarily an ill-defined one.

\begin{figure}
  \resizebox{8.5cm}{!}{\includegraphics{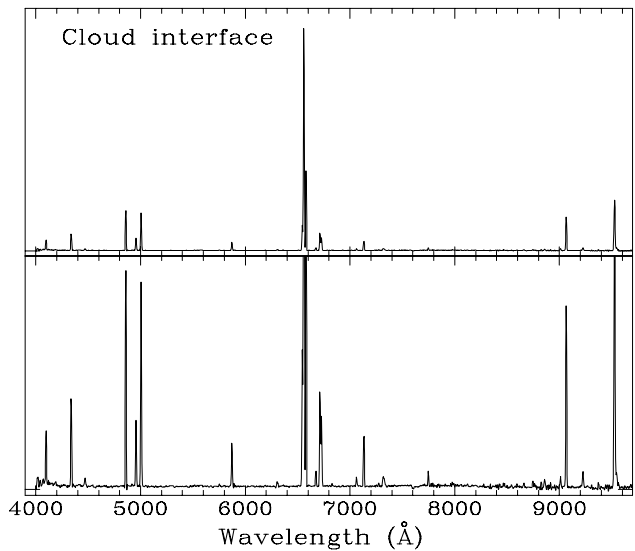}}
  \caption[]{Spectrum of the HII region at the interface where the
RCW~108 molecular clouds is being eroded by the radiation of
NGC~6193 and its brightest stars, HD~150135 and HD~150136. The
lower panel uses an expanded vertical scale to enhance the
visibility of the fainter emission lines.}
  \label{cloudinterface}
\end{figure}

\subsubsection{The cloud interface}

  The spectrum of the cloud interface, presented in
Figure~\ref{cloudinterface}, corresponds to that of a lightly
obscured nebula ($A_V \simeq 2.2$ from the H$\beta$/H$\alpha$
ratio) ionized by mid-to-late type stars. The [SII] line ratio
indicates a density somewhat below 100~cm$^{-3}$. The
extinction-corrected ratios of the [OIII]($\lambda 5007$), [SII]
($\lambda 6716$), [ArIII] ($\lambda 7138$), HeI ($\lambda 5876$),
and [SIII] ($\lambda 9531$) lines with respect to H$\alpha$ are
all compatible with an ionizing radiation temperature around
40,000~K. The strength of the [ArIII] and HeI features are not
consistent with temperatures significantly below that value, while
[SIII] does not allow for a much higher temperature and is in fact
better fitted by a temperature below 40,000~K (see
Section~\ref{hii_foreground}).

  The spectral type corresponding to a temperature of 40,000~K is
O7-O8, which must be compared to the spectral types directly
determined for HD~150135/150136, the two O-type stars of NGC~6193
that cause the ionization of the western cloud. This is in good
agreement with the spectral type O6/O7 given in the literature for
HD~150135. The spectral type of HD~150136 is often given as O5 in
the literature (e.g. Hiltner et al.~\cite{hiltner69}), which seems
to be earlier than suggested by the spectrum of the rim nebula.
The O7 classification given by Whiteoak~(\cite{whiteoak63}) for
this star is in better agreement with the spectrum of the nebula.

\begin{figure}
  \resizebox{8.5cm}{!}{\includegraphics{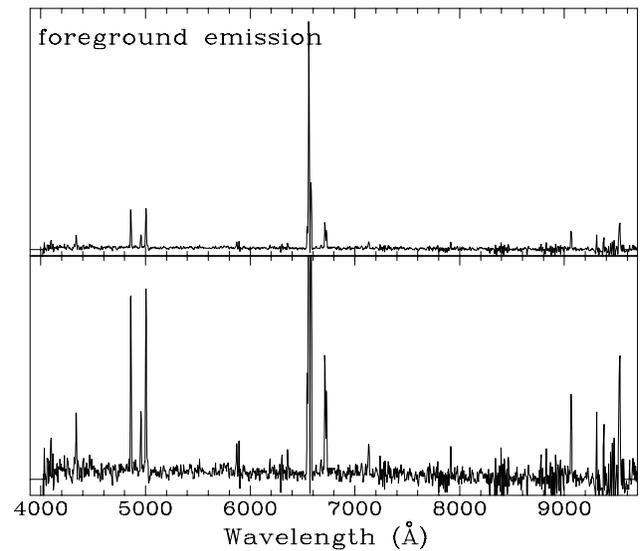}}
  \caption[]{Spectrum of the ionized gas foreground to the RCW~108
dark cloud. Like in Figure~\ref{cloudinterface}, the lower panel
uses an expanded vertical scale to enhance the visibility of the
fainter emission lines.}
  \label{foreground}
\end{figure}

\subsubsection{The foreground diffuse emission\label{hii_foreground}}

  The spectrum obtained through the part of the slit that runs
across the dark areas of the RCW~108 cloud still contains emission
lines similar to a low-intensity spectrum of the rim nebula,
indicating that the molecular cloud is seen through a layer of
ionized gas. The [SII]$(\lambda 6716) / (\lambda 6731)$ ratio is
lower than for the rim nebula, indicating a lower density that we
estimate at 10~cm$^{-3}$ or perhaps even less. The intensities of
the nebular lines are nearly constant across the slit with the
exception of a dark patch near the rim nebula where they have a
noticeably decrease, probably due to the existence of a cloud that
is embedded in this foreground layer and absorbs much of the
radiation from behind. We have used this local darkening to obtain
the sky spectrum to be subtracted from the spectrum of the
foreground emission. This sky-subtracted spectrum is shown in
Figure~\ref{foreground}.

  The spectrum of the foreground ionized gas is very similar to
that of the rim nebula, showing that the source of its ionization
is probably the HD~150135/150136 pair too. The extinction derived
from the H$\beta$/H$\alpha$ ratio, $A_V = 2.3$, is also very
similar to the extinction toward the rim nebula. However we notice
that the [SIII] lines in the far red are weaker here. The lower
temperature of the ionizing radiation that is inferred, around
30,000~K, is in turn incompatible with the line ratios of all the
other emission lines. Such a discrepancy, although less dramatic,
was also mentioned concerning the [SIII] lines of the rim nebula,
which also favored a temperature clearly below 40,000~K.

\begin{figure}
  \resizebox{8.5cm}{!}{\includegraphics{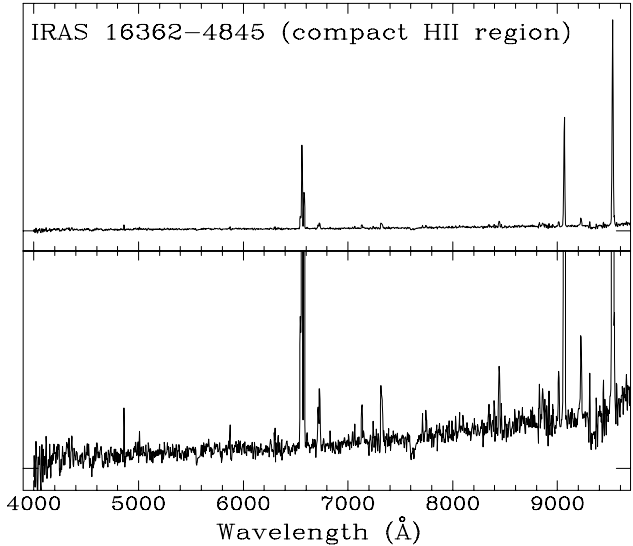}}
  \caption[]{Spectrum of the compact HII region. Like in
Figure~\ref{cloudinterface}, the lower panel uses an expanded
vertical scale to enhance the visibility of the fainter emission
lines.}
  \label{compact}
\end{figure}

\begin{figure}
  \resizebox{8.5cm}{!}{\includegraphics{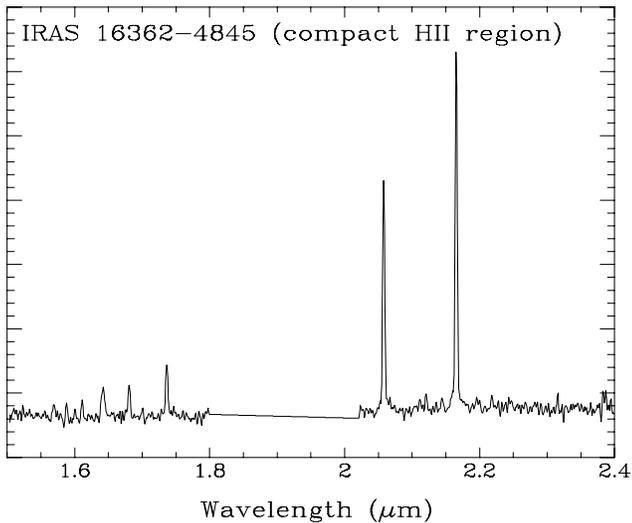}}
  \caption[]{Infrared spectrum of the compact HII region. The
interval between 1.8~$\mu$m and 2.02~$\mu$m has been removed due
to the strong telluric absorption.}
  \label{compactIR}
\end{figure}

\subsubsection{The compact HII region\label{hii_compact}}

  The extracted spectrum of the compact nebula, shown in
Figure~\ref{compact}, corresponds to a spot lying approximately
midway between the closest positions along the slit to stars~17
and 21 (see Figure~\ref{clustermap}), and is representative of the
whole compact component. For the background subtraction we have
chosen a region further to the East that contains the typical
spectrum of the foreground emission discussed in
Section~\ref{hii_foreground}, which as pointed out in that Section
keeps a fairly constant intensity across the area of the dark
nebula. Thus, we expect the spectrum presented in
Figure~\ref{compact} to be essentially uncontaminated by the
foreground emission.

  The most obvious differences between the visible spectrum of
the compact HII region and the ones of the rim nebula and the
foreground emission discussed earlier are due to extinction, for
which we obtain now a value between $A_V = 5.9$ (from
Paschen~6/H$\alpha$) and $A_V = 7.8$ (from H$\beta$/H$\alpha$).
The near infrared spectrum, shown in Figure~\ref{compactIR}, is
not extracted at exactly the same position as the visible one,
since the slit was oriented so as to simultaneously contain
stars~12 and 16, and corresponds to a spot about 5'' to the
Northwest of Star~12. It displays the features expected in a HII
region, with Brackett series lines clearly visible up to Br~10, as
well as a strong line of HeI at 2.058$\mu$m. The HeI(2.058) /
Br$\gamma$ ratio that we measure, 0.63, is near the lower edge of
the range measured in diverse HII regions. This might be a
consequence of the softness of the ionizing radiation, although
Lumsden et al.~(\cite{lumsden03}) caution against the use of this
ratio to derive the properties of the ionizing stars, especially
in compact HII regions.

  Extinction, together with the higher density of the emitting gas,
are the main responsible of the dominance of the [SIII] lines on
the far red part of the visible spectrum with respect to
H$\alpha$. The [SII] line ratio is markedly different now, and we
derive from it an electron density of 1,500~cm$^{-3}$. This is
much lower than the 13,800~cm$^{-3}$ independently derived from
radio continuum data by Urquhart et al.~(\cite{urquhart04}), which
may be due to the assumptions implicit in the derivation of
electron densities from radio continuum data, but also to the
possible existence of a denser, more deeply embedded component of
the HII region that does not dominate the visible emission
spectrum. Despite the high extinction the H$\beta$ and [OIII]
lines on the blue part of the spectrum are still visible. H$\beta$
is stronger than [OIII] now, arguing for a lower temperature of
the ionizing radiation than in the cases of the rim nebula and the
foreground emission. The lower temperature is also favored by the
other line ratios, particularly HeI/H$\alpha$ which has a good
sensitivity to temperature and only a mild dependency on density
in this range. The temperature of the ionizing radiation best
fitting all the line ratios is 35,000~K, and indicates a spectral
type later than those of HD~150135/150136. Indeed, the temperature
is in good agreement with a O9 spectral type, which in turn
matches well the position of Star~12 in the color-magnitude
diagram discussed in Section~\ref{aggregate}. The agreement with
the spectral type estimated by Urquhart et al.~(\cite{urquhart04})
based on the ionizing photon flux is also excellent.

\begin{figure}
  \resizebox{8.5cm}{!}{\includegraphics{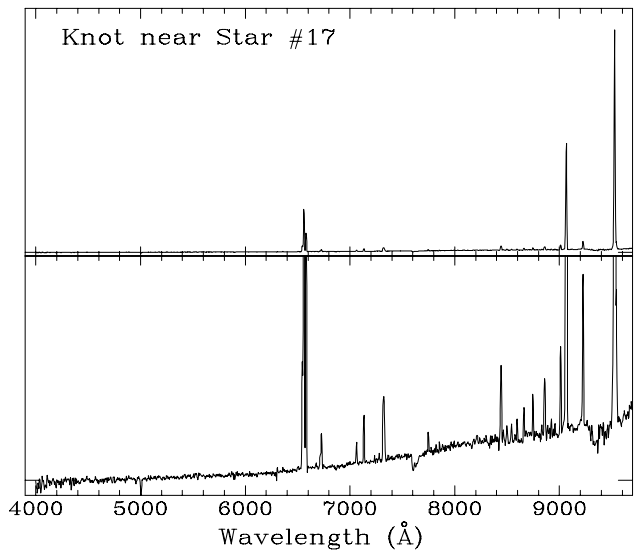}}
  \caption[]{Spectrum of the dense knot near Star~17. Like in
Figure~\ref{cloudinterface}, the lower panel uses an expanded
vertical scale to enhance the visibility of the fainter emission
lines.}
  \label{knot}
\end{figure}

\subsubsection{The emission knot near Star~17}

  The emission characteristics of the compact knot near Star~17
are similar to those of the compact HII region, with the main
differences of a stronger extinction ($A_V = 7.7$ from the
Paschen~6/H$\alpha$ ratio; the extinction below $\sim 6000$~\AA\
is too strong for H$\beta$ to be distinguishable). The $\lambda
6731$ component of the [SII] pair is now much stronger than the
$\lambda 6716$ component and the ratio between the two is close to
the high density saturation value, indicating a density in excess
of $10^4$~cm$^{-3}$. The spectrum longwards of H$\alpha$ is rich,
making the entire Paschen decrement between Paschen~6 and
Paschen~12 visible. Taking into account the difference in
densities inferred for the compact nebula and the knot, the
relative intensities of the [SII], [ArIII], and [SIII] emission
lines are all in agreement with the same ionizing radiation
temperature as derived for the compact nebula, $T_* \simeq
35,000$~K corresponding to a O9 spectral type. Given the proximity
of the knot to Star~12 and the consistency with its spectral type
as deduced from the spectrum of the compact nebula, we consider it
likely that it is a dense clump externally ionized by this star.

\subsection{Star formation across the RCW~108 molecular cloud\label{other_sf}}

  Our near infrared observations, covering a projected area of
nearly 5~pc~$\times$~5~pc, allow us to investigate the traces of
recent and ongoing star formation well beyond the immediate
vicinity of IRAS~16362-4845, at a scale that is intermediate
between the star formation directly associated to the compact HII
region and the large-scale, low resolution surveys of Yamaguchi et
al.~(\cite{yamaguchi99}) and Arnal et al.~(\cite{arnal03}) that
encompass the entire RCW~108 complex and most of the Ara~OB1
association. While IRAS~16362-4845 is an obvious site of current
massive star formation, other sites in the same cloud being
obscured by similar amounts and forming lower-mass stars may exist
as well without showing so obvious signposts. One of this sites is
CD-48~11039, already mentioned in Section~\ref{morphology}, which
is recognized by the reflection nebula that it illuminates. Other
young stellar objects of even lower mass may pass completely
unnoticed.

  To try to identify young lower mass stars in the region, we use the
position in the $(J-H)$, $(H-K_S)$ diagram as a telltale
diagnostic for the existence of hot circumstellar disks remnant
from their formation, a very common signature of youth among
intermediate-mass, pre-main sequence stars and a frequently used
approach to the identification of distributed star formation in
molecular clouds (e.g. Li et al.~\cite{li97}, Massi et
al.~\cite{massi00}, Brandner et al.~\cite{brandner01}, Jiang et
al.~\cite{jiang02}; see also Lada \& Lada~\cite{lada03} for a
review). To produce a reliable identification of near infrared
excesses we have considered only stars brighter than $K_S = 14.5$.
The luminosities to which this limit corresponds is obviously
highly dependent on the amount of light reprocessed by the
circumstellar material into the $K_S$ band and on how deeply
embedded in the cloud the star is. As a representative number, a
$K_S = 14.5$ star 1~Myr old having a foreground obscuration of
$A_V = 20$ ($A_K = 2.2$) as typically found in the IRAS~16362-4845
aggregate, and in which half of the emitted flux at $K_S$ comes
from reprocessed light would have a luminosity $L \sim
1.2$~L$_\odot$\footnote{We take a bolometric correction in the
$K_S$ band $BC_K = 2.1$ as a representative value for a 4,400~K
pre-main sequence star (Kenyon \& Hartmann~\cite{kenyon95}) in the
convective part of its evolutionary track (D'Antona \&
Mazitelli~\cite{dantona94})}, which corresponds to a mass of
0.8~M$_\odot$ at an age of 1~Myr.

  To assess the amount of infrared excess we use the
reddening-free quantity

$$Q = (J-H) - 1.70 (H-K_S) , $$

\noindent which measures the separation between the position in
the color-color diagram of a star with colors $(J-H)$, $(H-K_S)$,
and a reddening vector that traces the Rieke \&
Lebofsky~(\cite{rieke85}) extinction curve having its origin at
the intrinsic colors of a A0V star. Background red giants, which
align along a narrow strip running above this reddening vector in
the $(J-H)$, $(H-K_S)$ diagram, have $Q > 0$, while early-type
stars with no infrared excess cluster around $Q = 0$. Stars with
$Q < 0$ can be either late-type M dwarfs (Bessell \&
Brett~\cite{bessell88}) or stars with infrared excess. The former
are far too faint to be detected in our observations at the
distance of RCW~108, and the areal density of field late-M dwarfs
at the limiting magnitude $K_S = 14.5$ is too low for them to
appear in significant numbers in the imaged field (Reid et
al.~\cite{reid02}). A third possibility is that they may be
binaries unrelated to RCW~108 consisting of pairs with widely
different infrared colors, which should be a rare occurrence as
well.

\begin{figure*}
  \resizebox{18cm}{!}{\includegraphics{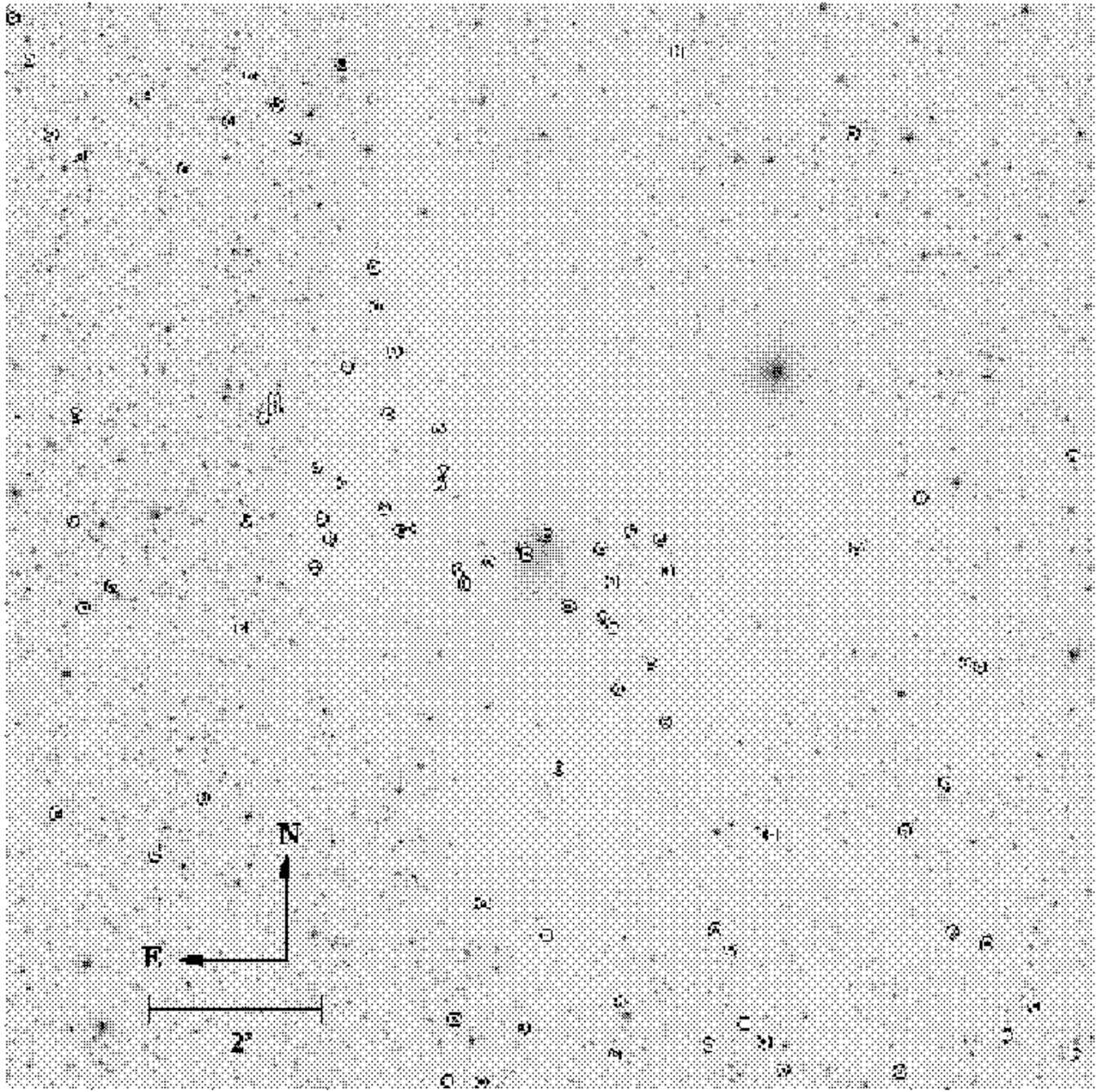}}
  \caption[]{$K_S$-band mosaic centered on the position of
IRAS~16362-4845, containing most of the RCW~108 cloud and parts of
its surrounding area. The circles mark the position of stars
brighter than $K_S = 14.5$ displaying infrared excess emission
according to the reddening-free $Q$ parameter defined in
Section~\ref{other_sf}.}
  \label{irexcess}
\end{figure*}

  Out of the 4,365 stars brighter than $K_S = 14.5$ for which
complete $JHK_S$ photometry is available, we have found 87
satisfying the $Q < -0.10$ criterion that we impose (rather
arbitrarily as far as the absolute value is concerned) as a
threshold defining the stars suspected to display infrared excess.
Their spatial distribution is plotted in Figure~\ref{irexcess},
superimposed on a $K_S$-band image of the region centered on
IRAS~16362-4845 and where the extent of the dense parts of the
RCW~108 clouds is well delineated by the areal density of stars.

  The distribution of stars with infrared excess in RCW~108 is
clearly non-random. The greatest concentration takes place in the
densest region of the cloud, near IRAS~16362-4845, where the
overall density of stars is lowest due to the presence of the
obscuring cloud. The infrared excess stars are not uniformly
scattered in the cloud, but are instead distributed along a belt
that contains IRAS~16362-4845 and runs from northeast to
southwest. The northeastern side is particularly interesting, as
it runs roughly parallel to the rim nebula in the inner side of
the cloud, suggesting that the stars in this region may have their
origin in the action of the ionization front encountering
molecular gas, perhaps producing the implosion of preexisting
cores by the increased external pressure. We note in particular a
clear contrast in density of near-infrared excess stars as we move
along an East-West strip immediately North of IRAS~16362-4845, in
which the presence of such objects ends about $1'$ East of the HII
region without any corresponding decrease in molecular gas column
density as we proceed further to the West. This suggests that,
despite the similar physical conditions of the molecular gas along
this strip, star formation has taken place only in the parts
closest to the ionization front. It is interesting to note that a
similar conclusion is independently reached by Urquhart et
al.~(\cite{urquhart04}) based on the location and expected
coevality of the three thermal sources detected in MSX
observations (one of which is IRAS~16362-4845), which they
classify as possible ultracompact HII regions corresponding to the
earliest stage of massive stellar evolution, at ages $<
10^5$~years. One of our infrared-excess stars in this zone, at
$\alpha(2000) = 16^h 40^m 11^s7$, $\delta(2000) = -48^\circ 48'
58''$, is surrounded by a tiny reflection nebula seen only in the
$J$-band image. Some stars with infrared excess are seen also
outside this area, where the ionization front has already
destroyed the cloud, and they may be the result of past episodes
of star formation when the cloud extended further towards
NGC~6193. Indeed, assuming that the freshly ionized gas can escape
freely leaving an essentially clear line of sight between the
ionizing stars and the rim nebula, taking our derived density of
$n_{H_2} = 3.9 \times 10^3$~cm$^{-3}$ for the molecular cloud and
assuming that the true distance $r$ from the ionizing stars to the
rim nebula is roughly the same as the projected distance, we
obtain a propagation velocity $v$ of the erosion front

$$ v = {{S_*} \over {4 \pi (2 n_{H_2}) r^2}} \simeq 5.6~{\rm km~s}^{-1}$$

\noindent where $S_* = 1.1 \times 10^{49}$~s$^{-1}$ is the flux
shortwards of the Lyman continuum of a O7 star (Schaerer \& de
Koter~\cite{schaerer97}). This is an upper limit to the velocity,
since the low density column of ionized gas between the stars and
the rim nebula must be kept ionized thus decreasing the ionizing
flux that reaches the molecular cloud, and the true distance
between the ionizing stars and the rim of the molecular cloud is
likely to be somewhat larger than the projected
distance\footnote{Based on their interpretation of infrared
emission produced in the photodissociation region at the rim
nebula/molecular cloud interface, Urquhart et
al.~(\cite{urquhart04}) suggest that NGC~6193 lies actually
somewhat behind the cloud.}. Nevertheless, the time needed for the
ionization front to travel at this speed from the position of the
easternmost infrared excess stars to the present position of the
rim of the cloud is only $\simeq 2.7 \times 10^5$~years (of the
same order of the estimate by Urquhart et al.~(\cite{urquhart04})
based on a different set of hypotheses), or over one order of
magnitude less than the typical disk dispersal timescales of
solar-type stars (Strom et al.~\cite{strom93}), so even a decrease
of the propagation velocity of the erosion front by the same
factor would still be consistent with the presence of stars with
infrared excess far beyond the present edge of the molecular
cloud. Overall, we find 39 stars (45\% of the total) on the side
of the cloud facing NGC~6193 and beyond the cloud in the same
direction, and thus possibly having a triggered origin.

  It is interesting to note however that numerous infrared excess
sources are found in the Southwestern quadrant of
Figure~\ref{irexcess}. The H$\alpha$ image
(Figure~\ref{wfi_halpha}) shows that this region is still fairly
opaque, despite the higher density of stars shown by the infrared
images. Also CO maps  such as those shown in Figure~\ref{chan_rcw}
show streamers of molecular gas extending in this direction. Based
on their location it seems more difficult to link the existence of
these stars, and in general all the stars located to the west of
IRAS~16362-4845, with the action of the HD~150135/150136 pair. No
other O-type stars have been identified near the Southwestern edge
of the RCW~108 cloud that may cause effects similar to those of
HD~150136/150136. The near-infrared stars near the Southwestern
edge of the cloud may thus represent star formation not being
triggered by an external cause. Evidence for the coexistence of
different modes (triggered/non-triggered) of star formation has
been also reported in other massive star formation regions (Jiang
et al.~\cite{jiang02}). However, the fact that such stars are also
non-randomly distributed across the cloud, and that there is no
distinct spatial separation between them and the stars for which
we proposed a triggered origin, puts a note of caution in our
interpretation of the latter, whose location on the side facing
NGC~6193 may also be fortuitous.

\subsubsection{Star formation efficiency in the western
cloud\label{efficiency}}

The derivation of the stellar mass of the RCW~108 complex on the
basis of the members identified via their infrared excess would
require determining their masses as well as the fraction of cloud
members that display such infrared excess. Unfortunately, the
currently available data prevent this exercise. However, it is
possible to use our estimate of the total mass of the cloud
(Section~\ref{masses}) together with indirect arguments to show
that the western cloud has in any case a low star formation
efficiency (defined as $SFE = M_* / (M_{\rm gas} + M_*)$, where
$M_*$ and $M_{\rm gas}$ are respectively the masses in stellar and
gaseous form) as compared to typical evolved giant molecular
clouds. The stellar mass estimate for the IRAS~16362-4849
aggregate, 210~M$_\odot$, represents only 2.6~\% of the mass in
molecular form (Section~\ref{masses}), well below the $\sim 10$~\%
of gas that is typically turned into stars in a giant molecular
cloud at the end of its life (Williams \&
McKee~\cite{williams97}). The fraction is even smaller if the
apparent deficiency of low mass stars in the aggregate hinted at
in Section~\ref{aggregate} is real. In order to raise the star
forming efficiency to 10~\% one thus needs to invoke the existence
of stars amounting to nearly 700~M$_\odot$ scattered across the
cloud outside the boundaries of the aggregate. If this component
obeyed a Miller-Scalo mass function, the expected number of stars
more massive than 3~M$_\odot$ would be around 50. Our observations
indicate that the number is far smaller than that, with
CD-48~11039 being the only star outside the aggregate with a mass
in this range. The low star-to-gas mass ratio, together with the
vast amount of molecular gas available to star formation and its
high volume density, thus suggest that most of star formation in
the RCW~108 still has to take place. It is interesting to
speculate that, should star formation proceed across RCW~108 in
the future until reaching a final $\sim 10$~\% efficiency, the
result may then be similar to the extended Orion Nebula Cluster.
The role played by IRAS~16362-4849 as the core of that future
RCW~108 aggregate would then be similar to the role that the
Trapezium plays in the Orion Nebula Cluster nowadays. In this
respect, RCW 108 and the Orion nebula complex might thus be
regarded as similar structures at different stages of their
evolution.

\section{Summary and conclusions}

  In this paper we have presented a collection of visible, near
infrared and millimeter observations of the RCW~108 region and its
associated rim nebula, focusing on its active massive star forming
site, IRAS~16362-4845, and on the molecular gas presumably
associated to the cluster NGC~6193. IRAS 16362-4845 is a compact
HII region whose actual morphology and stellar contents are best
revealed by near infrared observations. At the core of the HII
region lies a Trapezium-like compact cluster dominated by a late
O-type star, surrounded by a looser aggregate probably including
between one and three other late O-type stars and about 16 B-type
stars. The near-infrared color-magnitude diagram of the aggregate
gives an estimate of its mass of $\sim 210$~M$_\odot$. The cluster
seems to lack moderately reddened stars later than A0.

  We have discussed the spectrum of the ionized gas in the region
and the physical conditions of the molecular gas as inferred from
\co and \cod mapping. The compact HII region is dense ($\sim
1500$~cm$^{-3}$), although much less dense than the surrounding
molecular gas as inferred from the molecular-line observations in
the same direction. It spectrum suggests a spectral type O9 for
the ionizing star, in good agreement with the position of the
brightest star of its cluster in the color-magnitude diagram. A
compact knot located near the main ionizing star, whose density is
derived to be above $10^4$~cm~$^{-3}$ and whose lines indicate
ionization by a spectrum similar to that of the stars ionizing the
compact nebula, may thus be externally ionized by those same
stars. Streaming motions are revealed by the interferometric
H$\alpha$ observations both in the compact HII region and near the
molecular cloud interface, which can be interpreted in terms of
photoevaporation of the molecular gas.

  We obtain masses of 8000~M$_\odot$ and 660~M$_\odot$ for the
molecular clouds associated to RCW~108 and to NGC~6193,
respectively. The extinction on the background at the position of
IRAS~16362-4845 reaches up to $A_V = 70$~mag, while the peak
extinction produced by the cloud associated to NGC~6193 is much
lower, $A_V \simeq 18$. We consider explanations to the kinematics
of the NGC~6193 cloud based on rotation and on expansion powered
by the stars in the cluster, although we do not find conclusive
evidence for either interpretation. We attribute broad wing CO
emission (redshifted with regard to the bulk emission of the
molecular cloud) northwest of IRAS~16362-4845 to erosion of clumps
due to the ionizing gas of the comapct HII region. A more
complicated scenario including an expanding shell and outflow
emission from YSO can not be excluded but needs further
observational investigation.

  Using near infrared excess as a way to identify young stars
still surrounded by circumstellar dust, we find that their
distribution is inhomogeneous: not surprisingly, the molecular
cloud harbors most of them, mainly in the surroundings of
IRAS~16362-4845. Many are distributed in a broad band near the
edge of the molecular cloud and parallel to it, suggesting that
their formation may have been triggered by the progress of the
ionization front traced by the rim nebula into the molecular
cloud. Other infrared excess stars appear outside the cloud on the
side facing NGC~6193, suggesting that they formed in ancient
regions of the RCW~108 molecular cloud that have been eroded away.
However, numerous infrared excess sources appear also on the side
of the RCW~108 cloud opposite to NGC~6193, where there is no
obvious triggering candidate. Despite the evidence for widespread
ongoing star formation we find a low overall star formation
efficiency that, together with the physical conditions and mass of
the cloud, suggests that most of the star formation in RCW~108
still has to take place. Based on this, we speculate that
RCW~108 may represent a structure similar to the Orion nebula
complex observed at an earlier stage of evolution.

  The observations presented and discussed here, while not
being an exhaustive survey of the different components of RCW~108
and the gas associated to NGC~6193, provide some useful elements
towards a complete picture of a complex that features an emerged
cluster with massive stars and the last remnants of its parental
cloud, lying on the sky side-to-side to a young, embedded massive
star forming region in earliest stages of interaction between a
newly formed stellar aggregate and its surrounding gas. At a
larger scale, we can also investigate possible evidence of star
formation in a molecular cloud being triggered by an external
action, namely the ultraviolet radiation of the hottest stars of
NGC~6193. The identification of intermediate mass stars with near
infrared excesses across the cloud gives us a possibility of
carrying out follow-up observations aimed at deriving their
approximate ages, which might in turn provide evidence for a wave
of star formation running across the RCW~108 cloud. Finally, a
detailed study of the RCW~108 region can complement observations
at larger scales comprising the entire Ara~OB1 association, to
which the RCW~108 complex belongs, thus yielding so far elusive
observational views on the interplay of star forming processes at
widely different length scales.

\begin{acknowledgements}

  It is a pleasure to thank the staff of the La Silla Observatory for
  their support during our observations, especially Dr. Vanessa
  Doublier, Dr. Emanuela Pompei, Mr. Hern\'an N\'u\~nez, Mr. Duncan
  Castex, and Ms. M\'onica Castillo. We particularly thank Achim
  Tieftrunk for performing the molecular line observations using the
  SEST. We are also indebted to Dr.  Gra$\dot {\rm z}$yna
  Stasi\'nska for making available to us the results of her calculated
  spectra of model HII regions and for useful comments on their use,
  as well as to Dr. Margaret M. Hanson for her remarks on the infrared
  spectrum of the ionizing stars of IRAS~16362-4845. Useful
  discussions with Dr. Mario van den Ancker and Dr. Monika
  Petr-Gotzens on the nature of Star~16 are gratefully acknowledged.
  We thank Dr. P. Amram for the H$\alpha$ interferometric
  observations. Finally, we thank the anonymous referee for
  constructive comments that helped improving the presentation and
  contents of this paper.

\end{acknowledgements}

\end{document}